\documentclass[pre,aps,superscriptaddress]{revtex4-2}
\usepackage{amsfonts}
\usepackage{times}
\usepackage{amsmath}
\usepackage{graphicx}
\usepackage{amssymb}
\usepackage{xcolor}

\DeclareMathOperator{\sech}{sech}

\begin{document}

\title{Scattering-induced splitting of solitons in the discrete NLS equation with saturable nonlinearity}

\author{J.F. Tsoplefack}
\affiliation{Grupo de F\'{\i}sica No Lineal, Departamento de F\'{\i}sica Aplicada I,
Universidad de Sevilla. Escuela T\'{e}cnica Superior de Ingenier\'{\i}a Inform\'atica. Avda. Reina Mercedes s/n, 41012-Sevilla, Spain}
\affiliation{Institute of Nanoscience and Nanotechnology, National Center for Scientific Research “Demokritos”, Agia Paraskevi, 153 10 Athens, Greece}
\author{F. Palmero}
\affiliation{Grupo de F\'{\i}sica No Lineal, Departamento de F\'{\i}sica Aplicada I,
Universidad de Sevilla. Escuela T\'{e}cnica Superior de Ingenier\'{\i}a Inform\'atica. Avda. Reina Mercedes s/n, 41012-Sevilla, Spain}
\author{J. Cuevas-Maraver}
\affiliation{Grupo de F\'{\i}sica No Lineal, Departamento de F\'{\i}sica Aplicada I,
Universidad de Sevilla. Escuela Polit\'{e}cnica Superior, C/ Virgen de \'{A}%
frica, 7, 41011-Sevilla, Spain}
\affiliation{Instituto de Matem\'{a}ticas de la Universidad de Sevilla (IMUS). Edificio
Celestino Mutis. Avda. Reina Mercedes s/n, 41012-Sevilla, Spain}
\author{A. Provata}
\affiliation{Institute of Nanoscience and Nanotechnology, National Center for Scientific Research “Demokritos”, Agia Paraskevi, 153 10 Athens, Greece}
\author{D.J. Frantzeskakis}
\affiliation{Department of Physics, National and Kapodistrian University of Athens,
Panepistimiopolis, Zografos, Athens 15784, Greece}

\begin{abstract}

We study systematically the scattering of solitons on localized impurities in the discrete nonlinear Schr\"odinger (DNLS) equation with a saturable nonlinearity. We show that, apart from the generic scenario of the outcome of the scattering process, namely the emergence of a reflected and a transmitted soliton, other effects can occur.
In particular, it is found that, in the case of an attractive impurity, a soliton trapped at the impurity can coexist with the reflected and transmitted ones. This effect, which resembles the behaviour of a quantum particle interacting with a narrow impurity, has not previously reported for discrete setting. Parameter regimes are explored for determining soliton splitting on the impurity with special attention to equal soliton splitting.

\end{abstract}

\maketitle

\section{Introduction}

The study of nonlinear waves in discrete media has been a prolific subject of great interest for more than thirty years. Among such waves, one can highlight solitons, which are localized waves emerging in nonlinear dispersive media \cite{daux}. A relevant example concerns optical waveguide arrays (see, e.g., Ref.~\cite{kiva}), where the inherent periodicity produces discreteness, which in turn gives rise to dispersion. The latter, may be counterbalanced by nonlinear effects, as e.g., the Kerr effect (the dependence of the refractive index on the intensity of the light pulse). This leads to the emergence of discrete solitons or discrete breathers, depending on the context they emerge or the equation by which they are described \cite{daux,kiva,reviews_breathers}.

In the case of discrete optical settings, such as the one of waveguide arrays mentioned above, the electric field of light beams can be described, in the framework of the paraxial approximation, by the Discrete Nonlinear Schr\"odinger (DNLS) equation \cite{kiva,solitons_optics}. The properties of solitons in the DNLS equation depend strongly on the type of nonlinearity (which, at the same time, depends on the particular type of dielectric the waveguides consists of). The most ubiquitous nonlinearity is the cubic one, which appears not only in optical settings caused by the above mentioned Kerr effect (occurring, e.g., in AlGaAs), but also in Bose-Einstein condensates (BECs) trapped in deep optical lattices, granular crystals or biomolecules \cite{book_panos}. Other nonlinearities of interest are the cubic-quintic, power-law, or saturable (in the case of photorefractive materials, such as LiNbO$_3$ or SBN) ones  \cite{efremidis,morandotti,luther}. In this work, we focus on the last type of nonlinearity, and dub the DNLS equation with such nonlinearity as saturable DNLS or SDNLS equation, making comparisons with the analogously called cubic DNLS equation.

Contrary to the cubic DNLS equation, where the allowed frequencies for solitons are bounded only from below (with the lower bound corresponding to the linear limit), the frequencies available for SDNLS solitons have two bounds, with the upper one corresponding to the linear limit. For both equations, solitons become sharper when going beyond the linear limit. However, contrary to cubic solitons  whose width monotonically decreases with the frequency, the width of SDNLS solitons has a minimum when decreasing their  frequency, so that their width increases diverging when the frequency reaches the lower boundary \cite{maluckov}. Consequently, except for frequencies in the vicinity of the linear modes band, SDNLS solitons are in general wider than cubic ones. This is responsible, among other phenomena, for the possibility of radiationless motion of solitons in the SDNLS equation \cite{melvin_prl}, or the mobility of solitons in a two-dimensional saturable NLS lattice \cite{vicencio}, contrary to two-dimensional cubic NLS lattices where the generation of mobile solitons is impossible.

Having introduced one of the topics of this work, namely solitons in the SDNLS equation, we proceed by presenting the second one, i.e., the interaction of solitons with point impurities. This problem was already studied for the continuous NLS equation in the 1990s, with the works of Refs.~\cite{kosevich,Cao} (whose results were analytically proved and enlarged in Ref.~\cite{Goodman}) showing that, for attractive delta-like impurities, solitons can be trapped if their velocities are small and the impurity strength large enough, so that a nonlinear resonance with a defect mode occurs. In addition, this trapping can be supplemented with the splitting of the incoming soliton. The interaction with repulsive delta-like impurities shows a similar scenario but excluding the possibility of trapping \cite{Holmer}. In other words, the soliton behaves as a quantum particle which tunnels a potential barrier. This fact has been exploited in the generation of recombining solitons in harmonically trapped BECs \cite{Hulet}; such a generation process, has recently been demonstrated in experiments \cite{Wales}, and was shown to find applications in interferometry \cite{Polo,Sakaguchi}.
Notice, however, that this scenario becomes more complex with extended defects \cite{Ernst}. By virtue of Ehrenfest's theorem, the soliton can behave as a quantum particle if its length scale is larger than that of the defect \cite{Brand}. Consequently, narrow solitons (compared to the defect size), behave as classical particles and cannot split when interacting with defects \cite{Hansen}.

Once the main ingredients are presented, it is time to develop the main topic of the paper: the interaction of solitons with point impurities in the discrete NLS equation with saturable nonlinearity. Here it is important to point out that, despite the large number of studies devoted to the scattering of solitons from defects in continuous settings, there exist only a few works concerning discrete settings. Indeed, from the seminal work of Ref.~\cite{Bishop} where the dynamics of solitons on extended defects and point impurities was considered, very few examples can be found. For instance, in Ref.~\cite{Morales}, the trapping of low amplitude DNLS solitons by point impurities was considered, in Ref.~\cite{ricardo}
the scattering of highly localized discrete solitons from attractive and repulsive point impurities was analyzed, 
while in Ref.~\cite{Yaounde}, the interaction of multipeaked solitons with point impurities was studied. Surprisingly, all of these studies have been circumscribed to the cubic DNLS equation. To the best of our knowledge,
relevant studies in the framework of the SDNLS model have not been performed so far (even in continuum settings). The aim of the present work is to fill this gap.

Coming back to Ref.~\cite{ricardo}, it is relevant to briefly recopilate the observed phenomenology described therein. In particular, it was found that, when the impurity is repulsive, the soliton behaves as a classical particle interacting with a potential barrier, so that the soliton can ``cross'' the impurity if its speed is high enough (decelerating where the crossing occurs) and being reflected otherwise. For the case of an attractive impurity, a trapping of the soliton at the impurity site is mainly observed, creating a stationary state. If the impurity is strong enough, apart from the trapped breathing soliton, a reflected one emerges and, for very strong impurities, the trapped soliton cannot be generated. Finally, if the attractive impurity is weak, the soliton can cross it, while its velocity increases. In other words, the splitting of solitons has not been observed yet in DNLS settings.

However, for the case of the SDNLS equation that we study  herein, we find several differences. The most remarkable one is the possibility of soliton splitting for both attractive and repulsive impurities. In the former case, a soliton trapped at the impurity site can coexist simultaneously with the reflected and refracted solitons. We also observe that it is possible to find values of the velocity and the impurity strength for which the splinters possess the same power, although they can have different amplitude or width.

Our presentation is
organized as follows. In Section~\ref{sec:model} we introduce the model and some properties of stationary and moving discrete solitons of the saturable DNLS. The numerical results concerning the interaction between solitons and impurities are presented in Section~\ref{sec:numerics}. Finally, in Section~\ref{sec:conclusions}, we summarize the results of the study and present some ideas for further work.

\section{The model and its soliton solutions}
\label{sec:model}

\subsection{Model setup}

We consider
an array of coupled optical waveguides, filled with photorefractive dielectrics, which can be described by a 
SDNLS equation of the form:
\begin{equation}\label{eq:dyn}
    i\frac{d{\psi}_n}{dt}+C(\psi_{n+1}+\psi_{n-1}-2\psi_n)-\frac{\gamma}{1+|\psi_n|^2}\psi_n+\alpha_n\psi_n=0\,,\, n=-N/2,\ldots,N/2.
\end{equation}
Here, the wavefunction $\psi_n$ represents
the electric field envelope, $t$ denotes the propagation direction, $\gamma$ is the nonlinearity coefficient, $C$ the coupling constant, and $\alpha_n$ accounts for a linear inhomogeneity term. Experimentally, this kind of term can be implemented, e.g., by varying the separation between
a single pair of adjacent waveguides in an homogeneous array. In what follows, we consider this term to be of the form $\alpha_n=\alpha\delta_{n,0}$, i.e., it describes a single point defect in the lattice that can be either attractive ($\alpha>0$) or repulsive ($\alpha<0$). We consider here the case of a focusing nonlinearity, with $\gamma>0$ (and a positive coupling constant $C$); upon suitable parameter renormalization, we will fix the value of $\gamma$, without lack of generality, to $\gamma=1$.

The SDNLS equation possesses two conserved quantities, namely the power (squared $\ell^2$ norm):
\begin{equation}\label{eq:norm}
    P=\sum_n|\psi_n|^2,
\end{equation}
and the Hamiltonian:
\begin{equation}\label{eq:hamiltonian}
    H=\sum_n\left[C|\psi_{n+1}-\psi_n|^2+\gamma\log(1+|\psi|^2)-\alpha_n|\psi_n|^2\right].
\end{equation}

\subsection{Stationary and moving solitons}

Stationary solutions of equation (\ref{eq:dyn}) can be sought in the form:
\begin{equation}
    \psi_n(t)=\phi_n\exp(-i\omega t)\,,
\end{equation}
with $\omega$ being the soliton frequency. With this assumption, (\ref{eq:dyn}) transforms into an algebraic set of equations:
\begin{equation}\label{eq:stat}
    \omega\phi_n+C(\phi_{n+1}+\phi_{n-1}-2\phi_n)-\frac{\gamma}{1+|\phi_n|^2}\phi_n+\alpha_n\phi_n=0.
\end{equation}
Two important quantities to account for are the soliton's center of mass ($X$) and width ($W$) defined as follows:
\begin{equation}\label{eq:XW}
    X=\frac{1}{P}\sum_n n|\psi_n|^2, \qquad W=\sqrt{\frac{1}{P}\sum_n n^2|\psi_n|^2-X^2}.
\end{equation}

\begin{figure}
\begin{tabular}{cc}
\includegraphics[width=6cm]{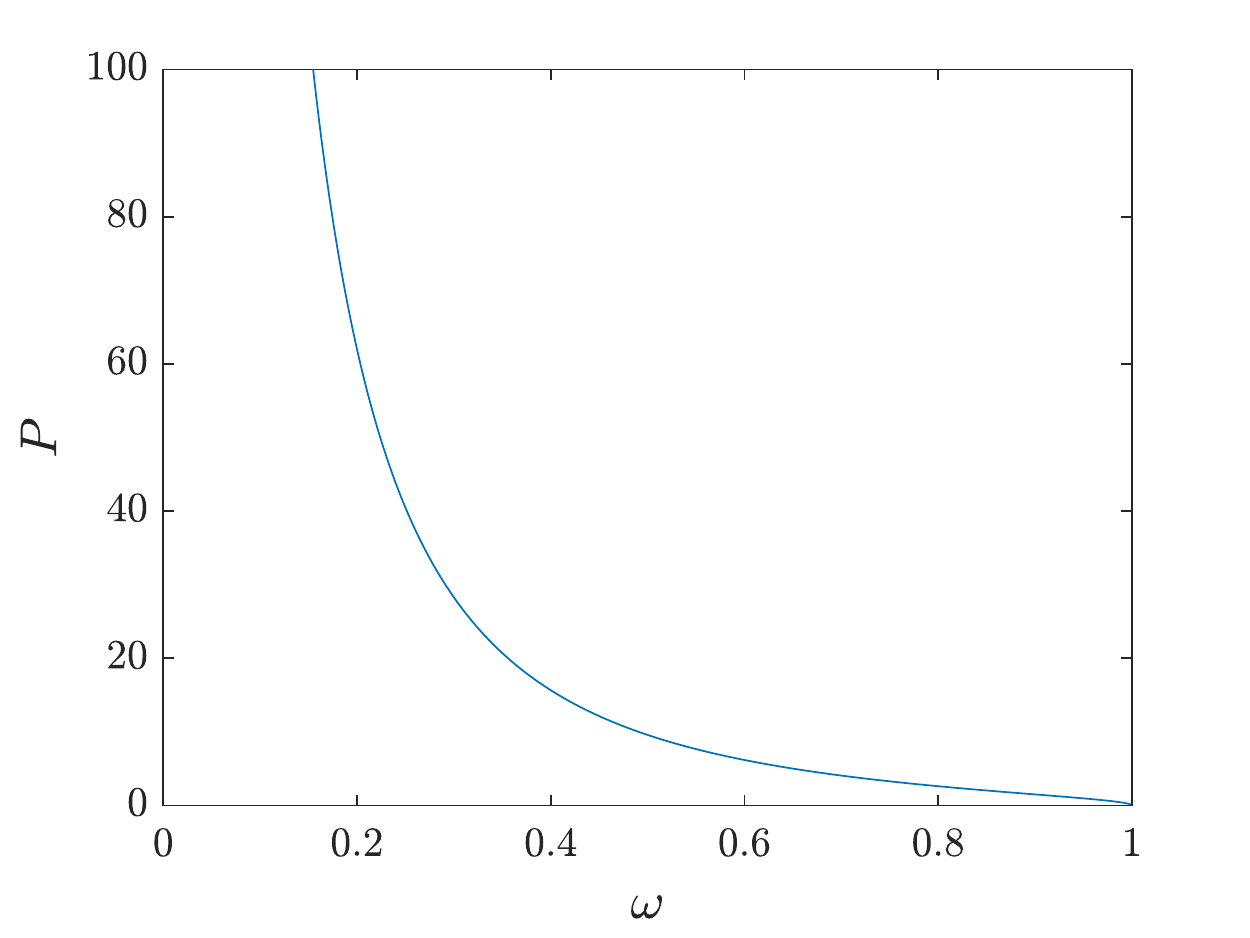} &
\includegraphics[width=6cm]{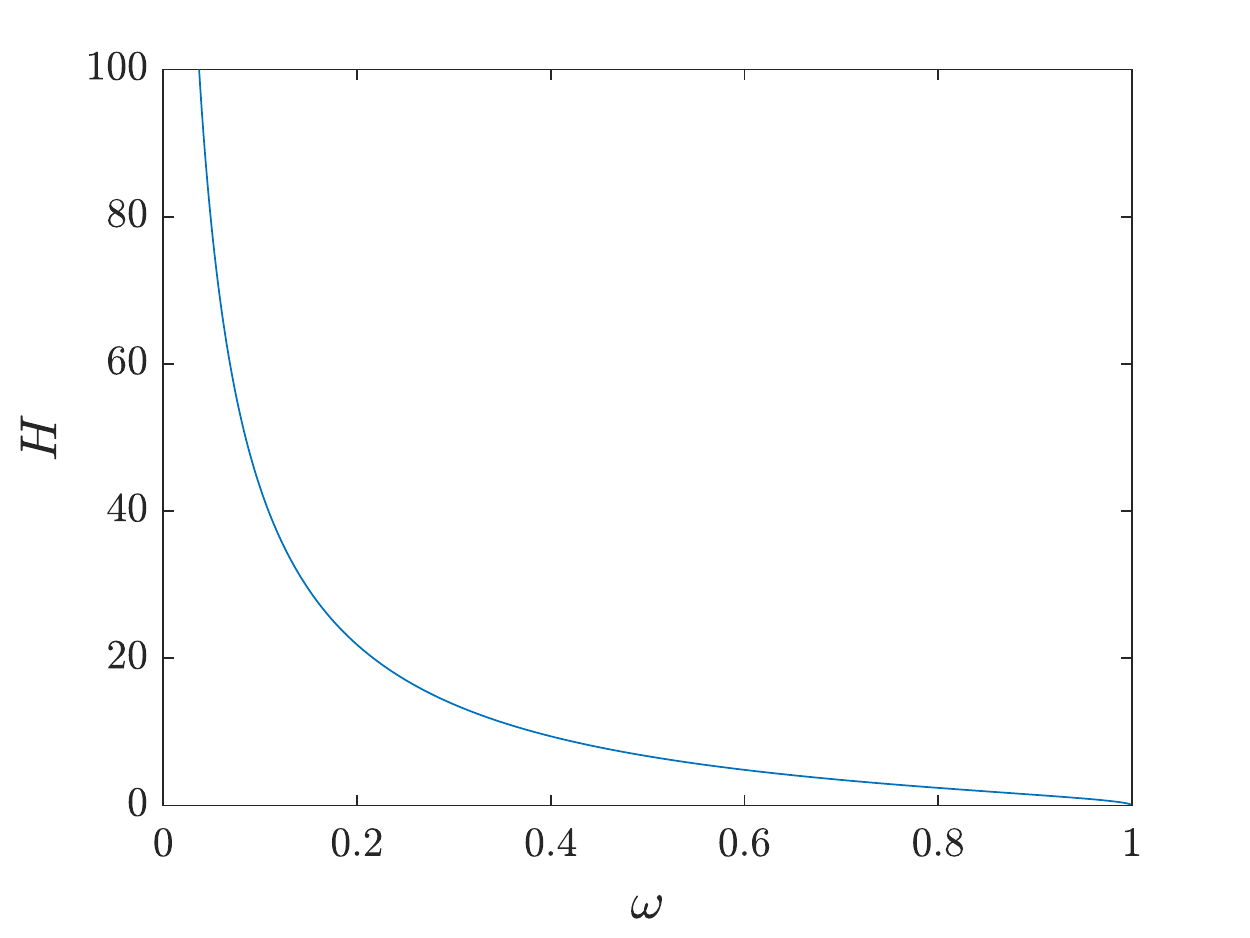} \\
\includegraphics[width=6cm]{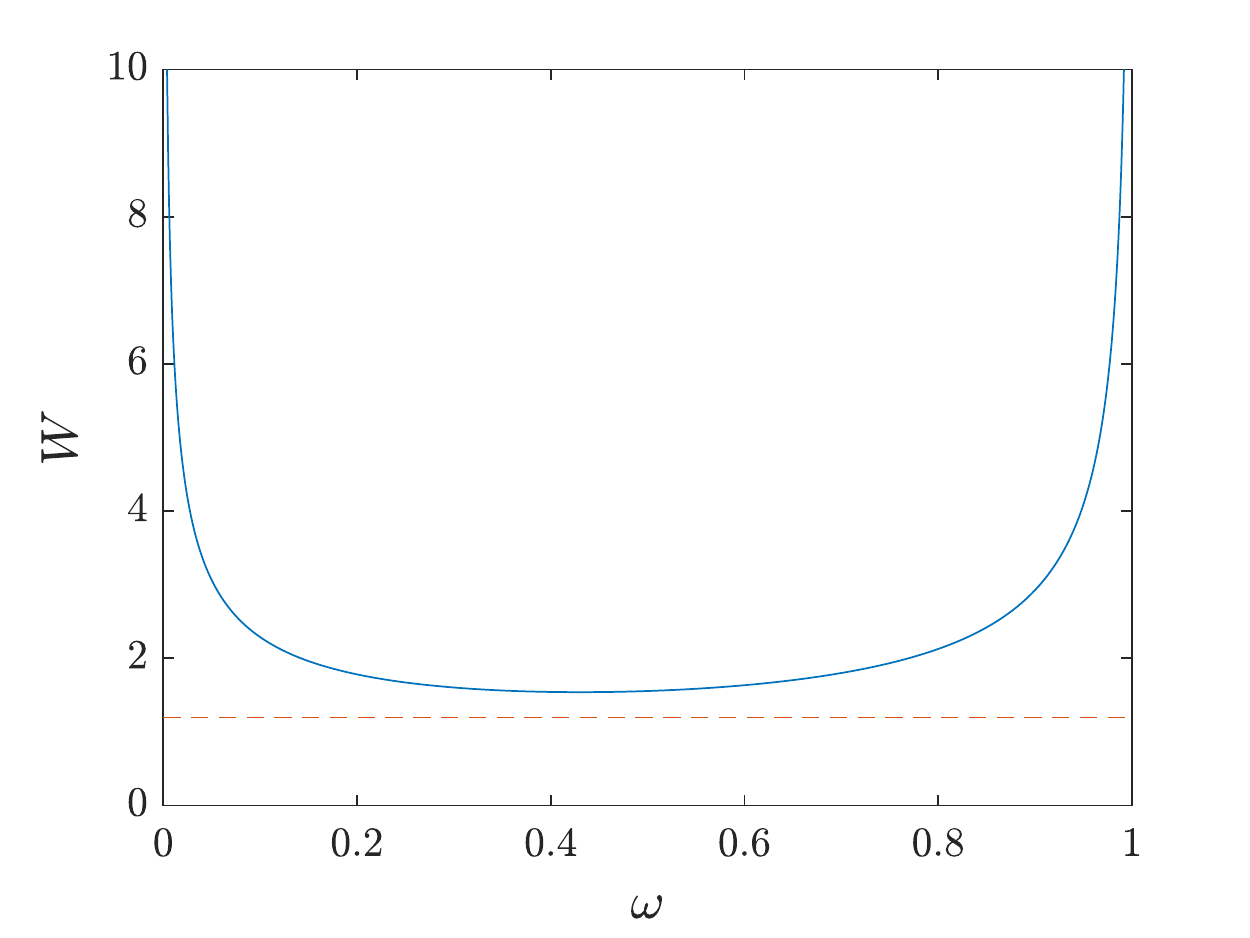} &
\includegraphics[width=6cm]{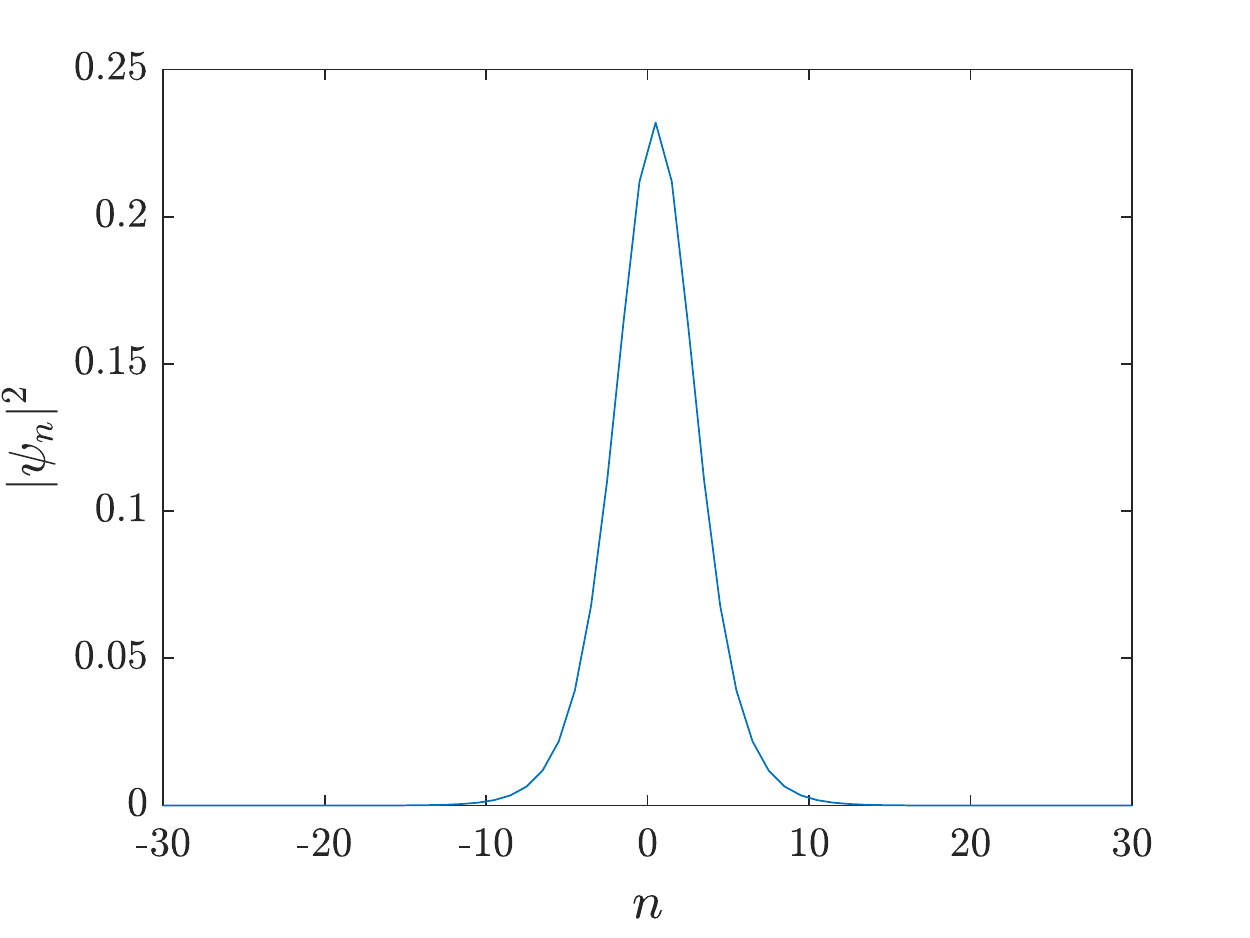}
\end{tabular}%
\caption{Frequency dependence
of the power (top left panel), energy (top right panel) and width (bottom left panel) of solitons in a  homogeneous lattice ($\alpha_n=0$) with $C=1$. In the
bottom left panel, the dashed (red) line 
corresponds to the width of the cubic DNLS soliton considered in the work of \cite{ricardo}. Bottom right panel depicts the profile of the soliton density, $|\psi_n|^2$, for $\omega=0.9$ (the moving soliton profile is, obviously, the same).}
\label{fig:stationary}
\end{figure}

We will be interested in generating moving solitons far from the defect site that are launched towards the latter. To this aim, first we need to calculate a stationary soliton centered at $n=n_0$, which is solution of Eq.~(\ref{eq:stat}) with $\alpha_n=0$; this equation is solved by means of a fixed-point algorithm (e.g., Newton-Raphson) using a quasi-continuum seed, $\phi_n=\sech(n-n_0)$ for $\omega=0.9$ and $C=1$. Once the stationary soliton for such values of $\omega$ and $C$ is found,
it can be continued to the desired values of those parameters. Figure~\ref{fig:stationary} shows the frequency dependence of the power, energy (Hamiltonian) and width of solitons for $\alpha_n=0$ and $C=1$; note that this value of the coupling constant will be fixed for the rest of the paper for reasons explained below. It can be observed that $0<\omega<\gamma$ and, since $\gamma=1$ (as stated in the previous subsection), $\omega\in(0,1)$, with the energy and power tending to zero when $\omega\rightarrow1$ and diverging when $\omega\rightarrow0$. As pointed out in the Introduction, the frequency is bounded, contrary to the cubic DNLS equation where the frequency takes values in a semi-infinite interval. Regarding the width, we can observe that it diverges in the limits $\omega\rightarrow0$ and $\omega\rightarrow1$, having a minimum at $\omega=0.43$; this behaviour is in stark contrast with the one observed in the cubic DNLS, where the width decreases when the frequency departs from the linear modes band. For the sake of comparison,  Fig.~\ref{fig:stationary} shows a dashed (red) line indicating the width of the soliton considered in \cite{ricardo}, which is smaller than that of any SDNLS soliton.

Once a stationary soliton is attained (see Fig.~\ref{fig:stationary} for an example of its profile), it is kicked with a thrust $q$ in the following form:
\begin{equation}
    \psi_n(0)=\phi_n\mathrm{e}^{iq(n-n_0)}.
\end{equation}
When this initial condition is used for simulating Eq.~(\ref{eq:dyn}), a moving soliton can be generated, as long as the Peierls-Nabarro barrier is small enough. This can only be attained when $C$ is large enough, although, as explained in \cite{melvin_prl}, this does not warrant the mobility. Nevertheless, we have checked that the solitons can be set into motion emitting a low amount of radiation for $C=1$ and $\omega=0.9$ whenever $q\in[0.05,0.5]$. Because of this, we implement such a restriction in $q$ in what follows. As the free soliton can be considered as a classical  quasi-particle of effective mass equal to $P$, and $\phi_n\in\mathbb{R}^n$, the velocity $v=dX/dt$ is related to the thrust $q$ via
\begin{equation}\label{eq:velocity}
    v=\frac{2\sin(q)}{P}\sum_n\phi_{n+1}\phi_n.
\end{equation}
For our particular case, $v=1.9687\sin(q)$. From this relation, it is easy to find the kinetic energy of the quasi-particle as $K=5.9584\sin^2(q/2)$.

\section{Interaction of solitons with impurities} \label{sec:numerics}

In what follows we consider solitons with frequency $\omega=0.9$ and coupling constant $C=1$, which are launched from $n_0=-50$ (unless stated otherwise) in a lattice with $N$ nodes and periodic boundary conditions, and a delta-like impurity of strength $\alpha$ is located at $n=0$; the number $N$ has been taken large enough so that the boundaries do not affect the outcome. In order to monitor the outcome, we define the reflection ($R$), transmission ($T$), and trapping ($L$) coefficients as follows:
\begin{equation}\label{eq:coefs}
    R=\lim_{t\rightarrow\infty}\frac{1}{P}\sum_{n=-N/2}^{-(\delta+1)}|\psi_n(t)|^2\,,\quad
    T=\lim_{t\rightarrow\infty}\frac{1}{P}\sum_{n=\delta+1}^{N/2}|\psi_n(t)|^2\,, \quad
    L=\lim_{t\rightarrow\infty}\frac{1}{P}\sum_{n=-\delta}^{\delta}|\psi_n(t)|^2\,,
\end{equation}
where $P$ is soliton power [see Eq.~(\ref{eq:norm})] and $\delta=20$ (i.e. we are assuming a typical trapped soliton width of $\lesssim 41$ nodes, cf.  Fig.~\ref{fig:simsplit_attractive} below). The limit
$t\rightarrow\infty$ in the above definitions has the following meaning: the quantities $R$, $T$ and $L$ are determined at values of $t$
large enough
after the collision event but, at the same time, small enough for neglecting the interaction between solitons via the periodic boundaries.

\begin{figure}
\begin{tabular}{cc}
\includegraphics[width=6cm]{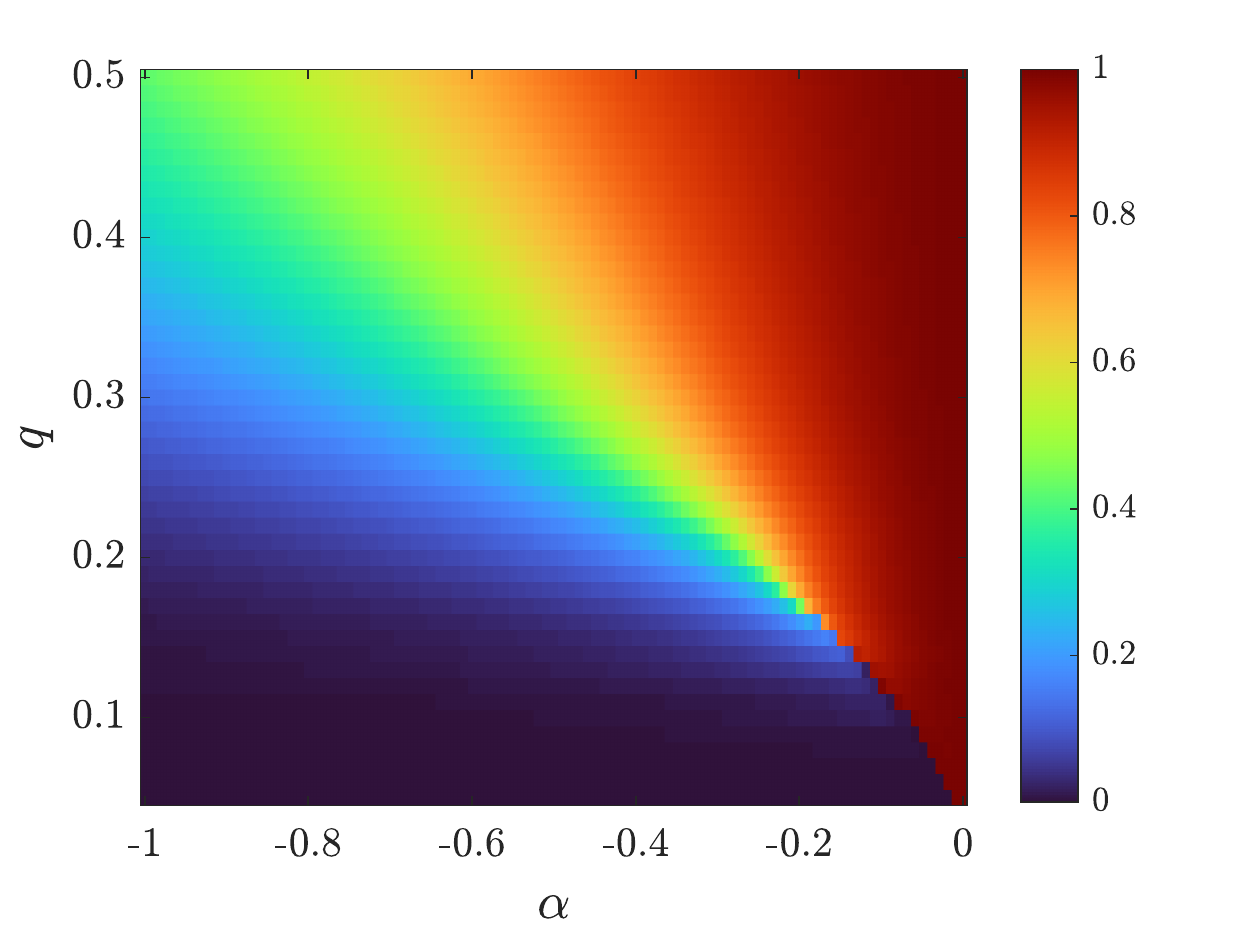} &
\includegraphics[width=6cm]{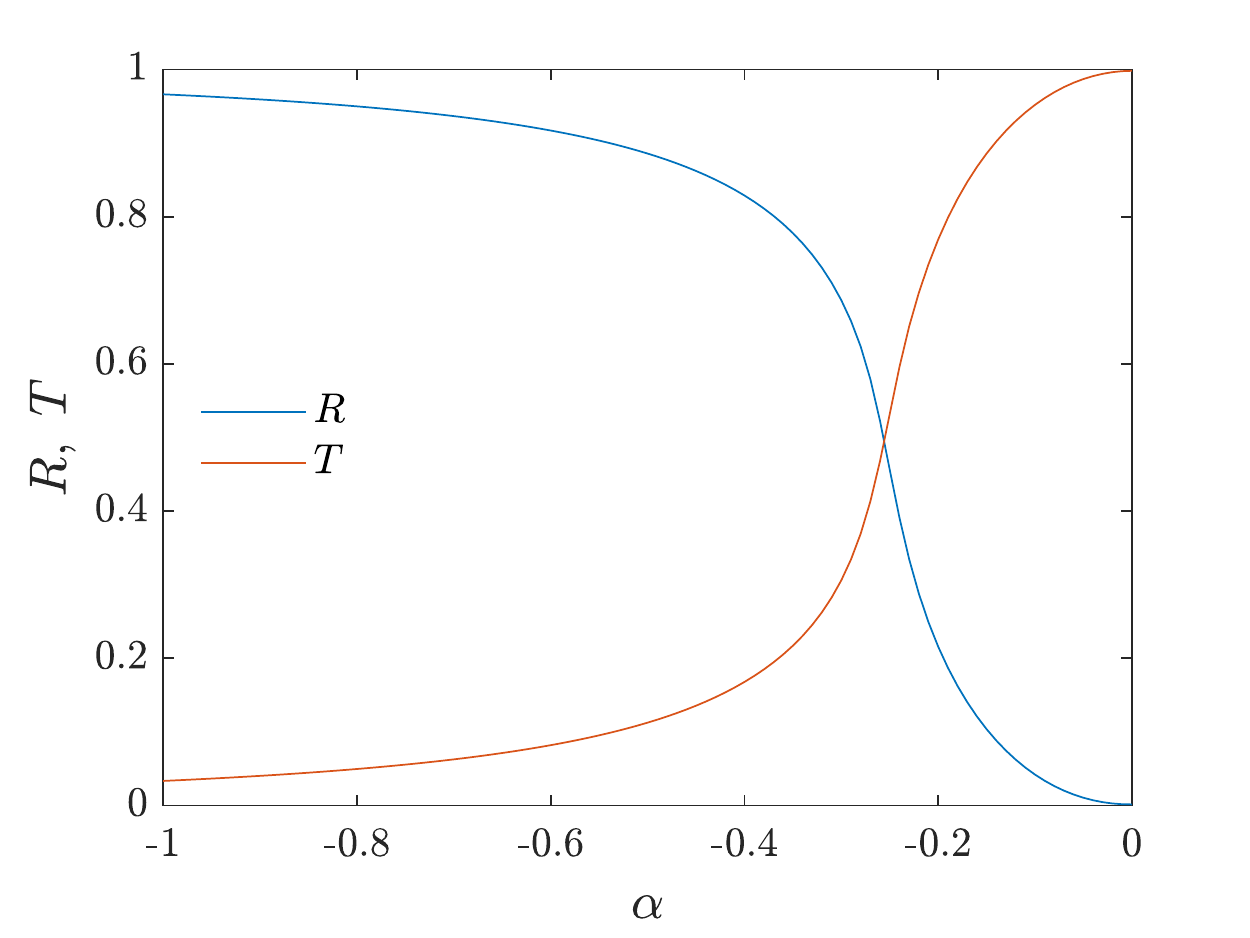} \\
\end{tabular}%
\caption{Repulsive impurity. Left panel: dependence of $T$ on $\alpha$ and $q$. Right panel: dependence of $R$ and $T$ on  $\alpha$ for $q=0.2$. In all the simulations, $N=1000$.
}
\label{fig:TR_repulsive}
\end{figure}

\begin{figure}
\begin{tabular}{cc}
\includegraphics[width=6cm]{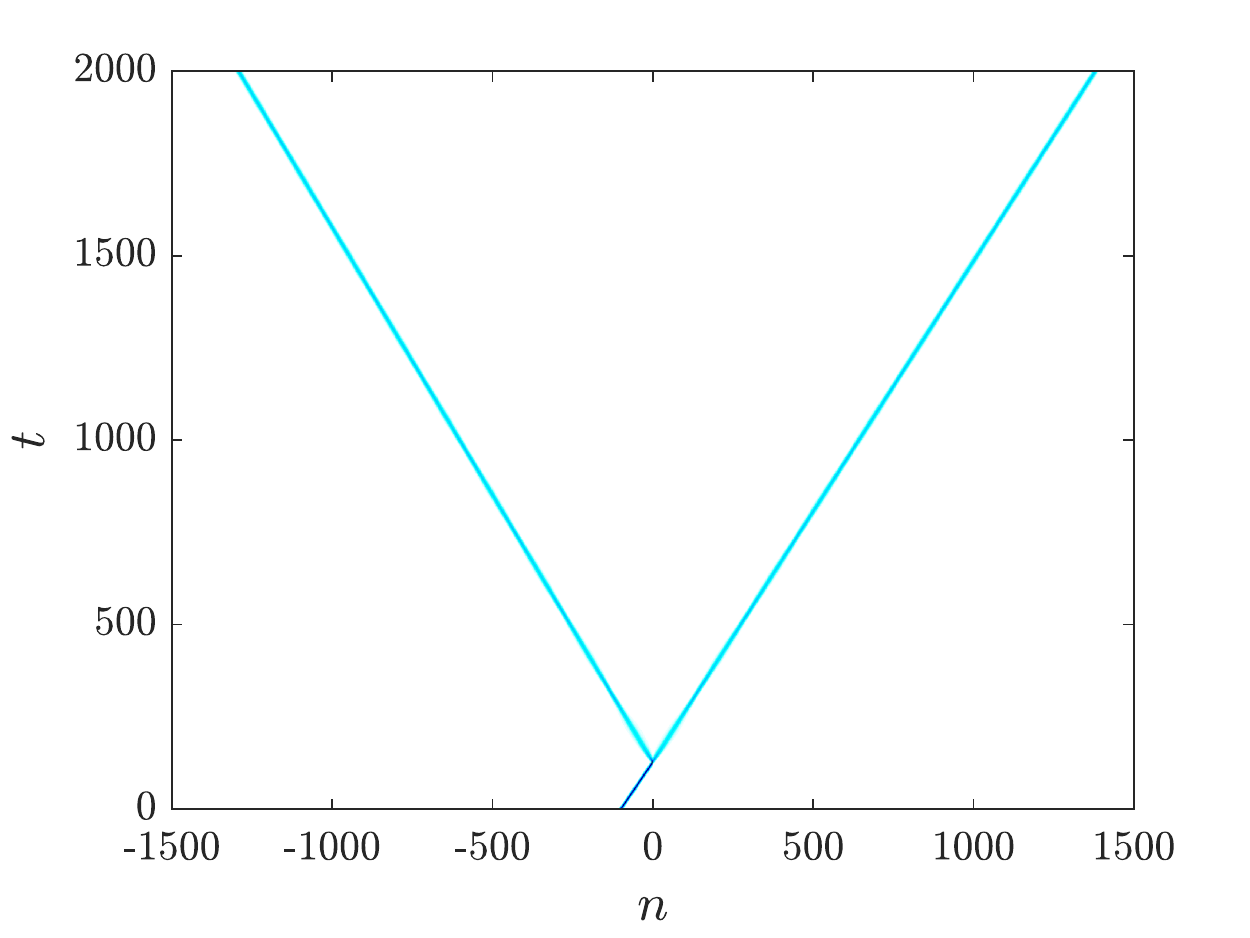} &
\includegraphics[width=6cm]{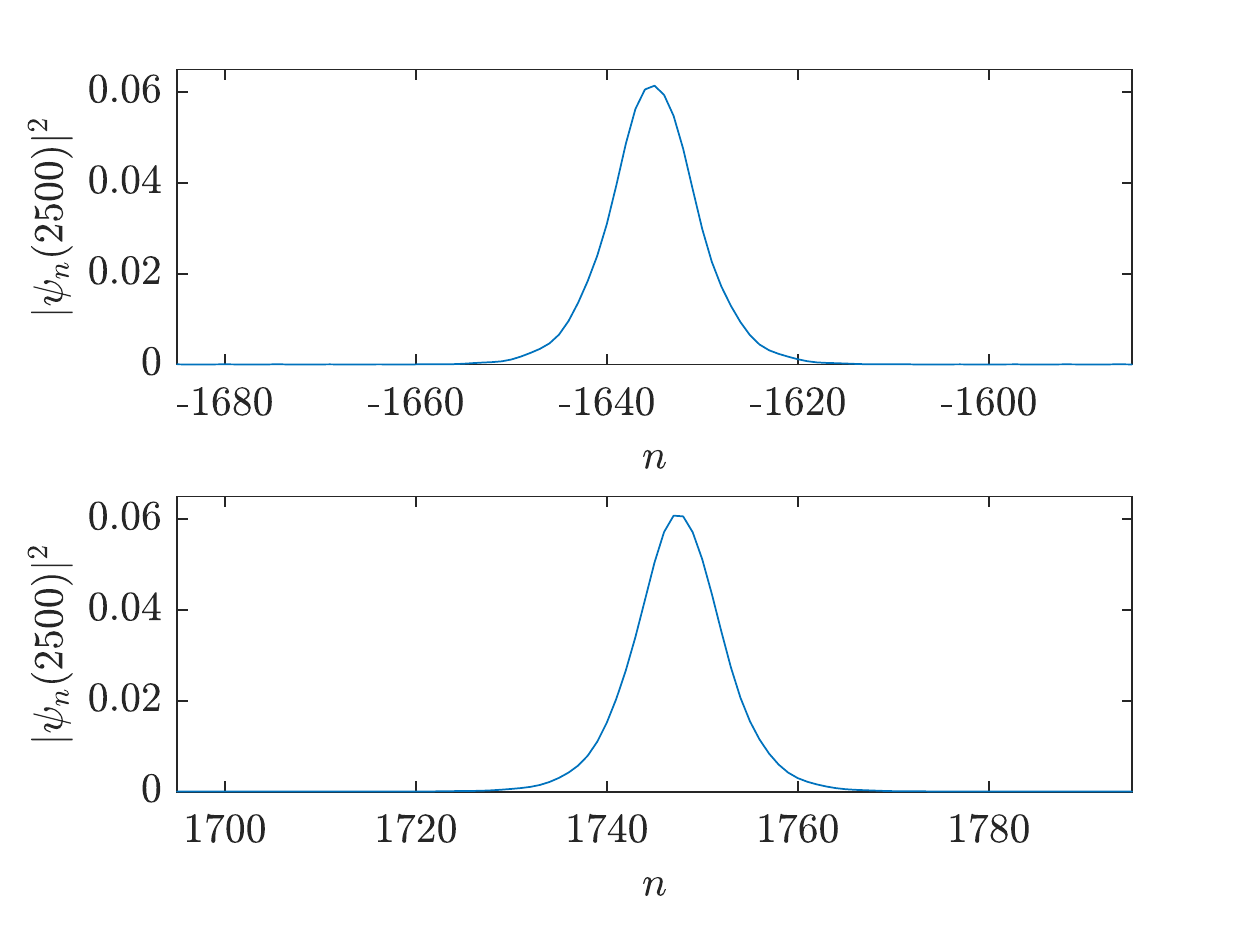} \\
\end{tabular}%
\caption{Evolution of a soliton with $q=0.41$ interacting with a repulsive impurity with $\alpha=-0.70$, which results in equal splitting. Left panel shows the space-time diagram of the density $|\psi_n|^2$, and right panel shows the profiles of the splinters a long time ($t=2500$).}
\label{fig:simsplit_repulsive}
\end{figure}

First, we consider the case of repulsive impurity, i.e., $\alpha<0$. In this case, the outcome is quite simple: the soliton breaks into a reflected and a transmitted soliton, similarly to the scattering of a quantum particle from a potential barrier. As is expected, $T$ decreases with $|\alpha|$ and, at the same time, $R$ grows with $|\alpha|$. Figure~\ref{fig:TR_repulsive} shows the dependence of $T$ on $\alpha$ and $q$ ($R$ is simply equal to $1-T$). Consequently, there is a critical value of $\alpha$, namely $\alpha_c$, depending on $q$, for which equal splitting takes place, i.e. $T=R=0.5$. This equal splitting
results in the generation of two solitons with the same power, although it does not mean that the solitons have the same width, amplitude or velocity. Specifically, as can be seen in the example of Fig.~\ref{fig:simsplit_repulsive} depicting the evolution of a soliton with $\alpha=-0.7$ and $q=0.41$ (corresponding to a velocity $v=0.785$), the splinters move with velocities $0.737$ and $-0.690$ in spite that $R=T=0.5$.


We can compare this outcome with relevant phenomenology reported previously in the literature. In particular, the work of \cite{Holmer} considers the scattering of a soliton in the NLS from a Dirac delta potential obtaining that, if the soliton velocity $V$ is large, the transmission coefficient is the same as that of a quantum particle in the linear Schr\"odinger equation $i\partial_t\psi=[-(1/2)\partial_{x}^2+\alpha\delta(x)]\psi$, namely:
\begin{equation}\label{eq:T_linear}
    T(V)=\frac{V^2}{V^2+\alpha^2},
\end{equation}
so a perfect splitting would take place at $|\alpha|=|\alpha_c|$ with $|\alpha_c|=V$. In our case, the DNLS equation (\ref{eq:dyn}) can be considered as a finite differences discretization of the NLS equation, with the 2nd-order spatial derivative being
$(1/2)\partial_{x}^2\psi\approx(1/2h^2)(\psi_{n+1}+\psi_{n-1}-2\psi_n)$. Upon comparing this expression with the discrete Laplacian term of (\ref{eq:dyn}), we get that $C=1/2h^2$. As $x=nh$ the discrete soliton velocity $v$ is related to that of the continuum soliton by $v=V/h$, or in terms of the coupling constant, $v=V\sqrt{2C}$.

\begin{figure}
\begin{tabular}{cc}
\includegraphics[width=6cm]{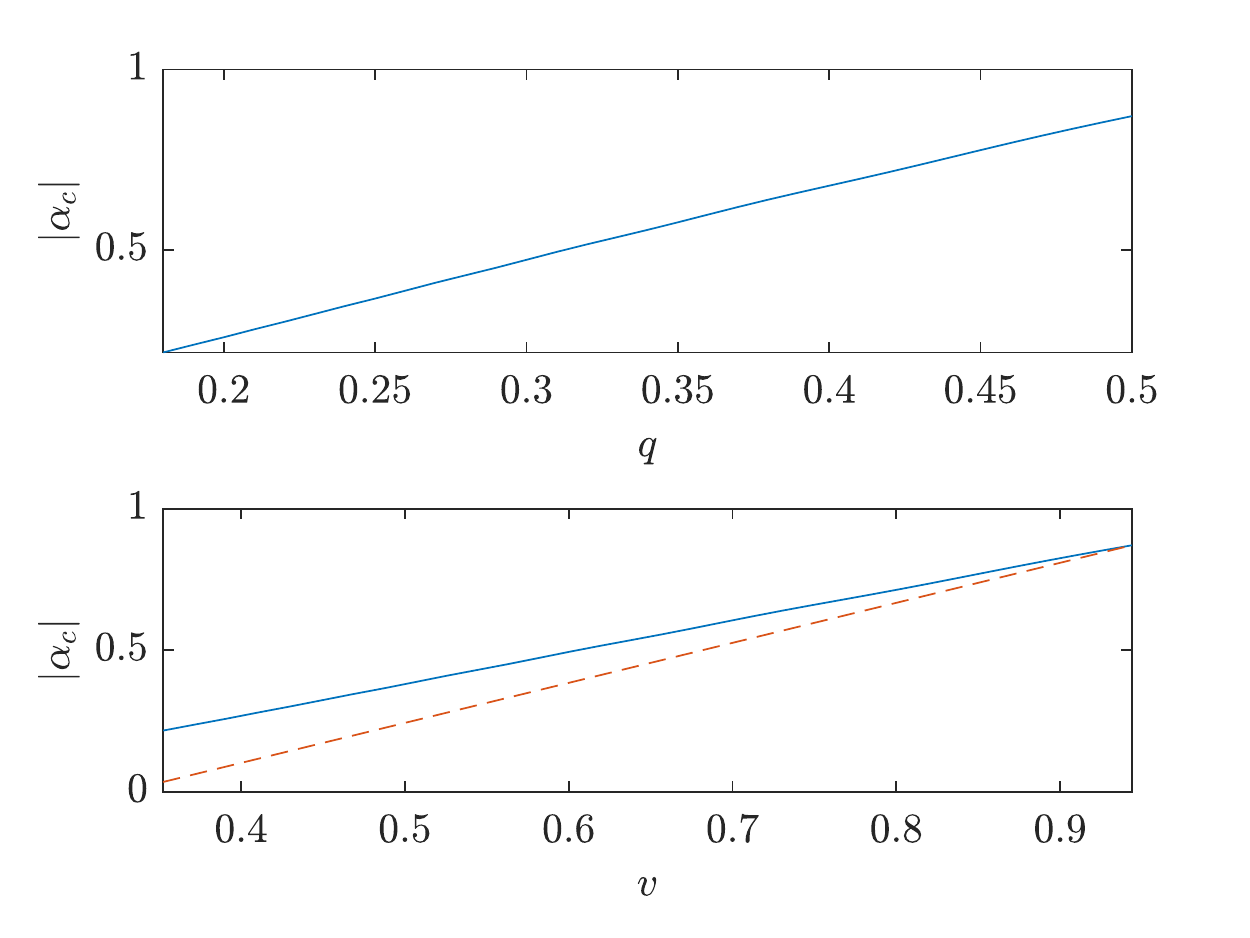} &
\includegraphics[width=6cm]{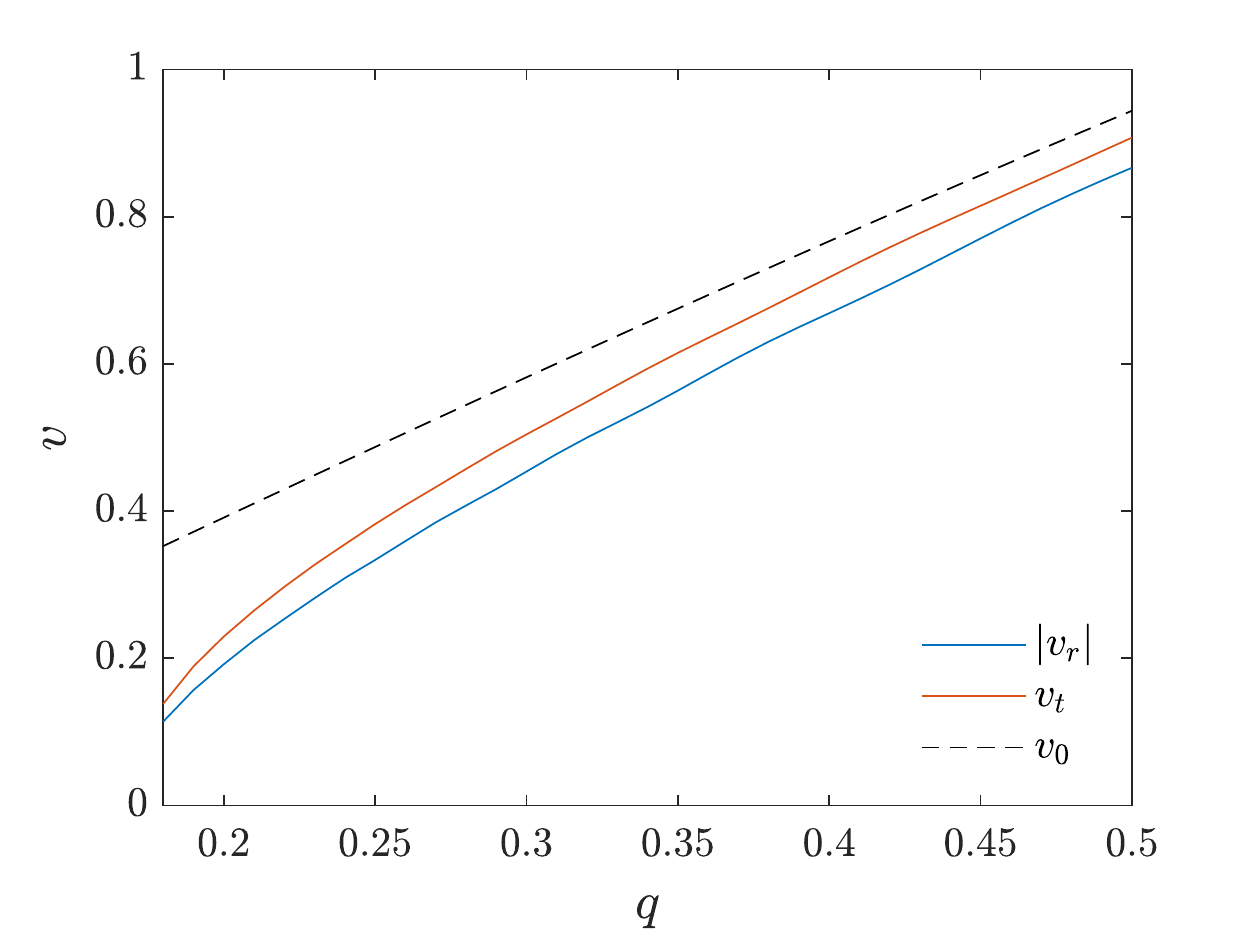} \\
\end{tabular}%
\caption{Analysis of equal splitting for repulsive impurities. Left panel shows the value of $|\alpha_c|$ versus $q$ and $v$; the (red) dashed line in the bottom plot corresponds to a line with slope $\sqrt{2C}$, corresponding to the theoretical prediction of the NLS equation. Right panel shows the velocities of the transmitted $v_t$ and reflected $v_r$ splinters together with the velocity of the incoming soliton.}
\label{fig:split_repulsive}
\end{figure}

With this in mind, one can see that in the discrete case, $|\alpha_c|$ should tend to $v\sqrt{2C}$ for fast enough solitons if the scattering behaves for the DNLS in a similar fashion to the NLS case. We can check that this is the case in Fig.~\ref{fig:split_repulsive} which, apart from $\alpha_c(q)$ and $\alpha_c(v)$, shows the velocities of the splinters as a function of $q(\alpha_c)$ in the case  $R=T=0.5$. From the latter figure, it can be seen that the velocity of the transmitted splinter is always higher than that of the reflected one, so that the scattering process seems to have a preference to ``remember'' the direction of the incoming soliton. In addition, one can see that the velocities of the splinters approach to the velocity of the incoming soliton when the latter increases.
%

\begin{figure}
\begin{tabular}{ccc}
\includegraphics[width=6cm]{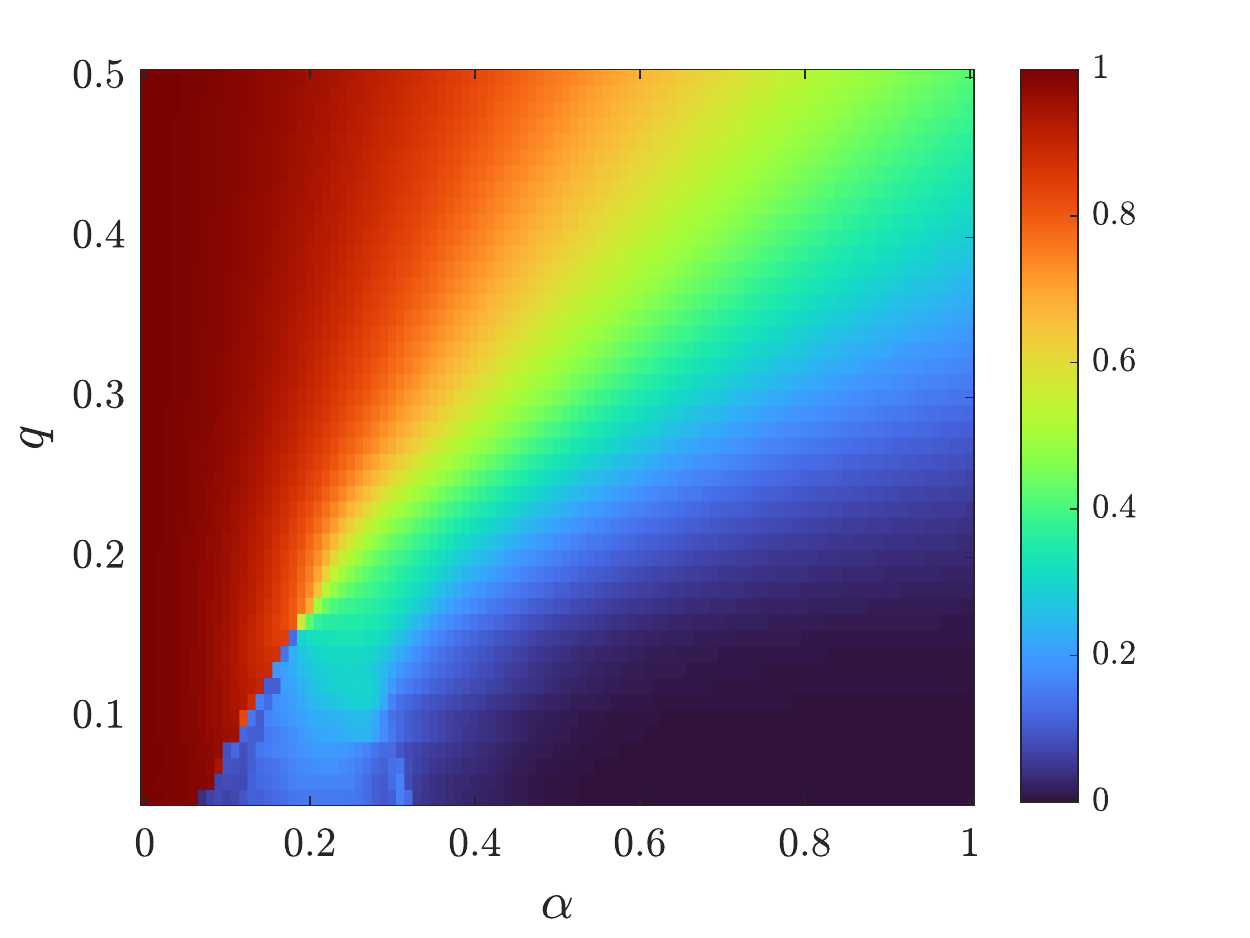} &
\includegraphics[width=6cm]{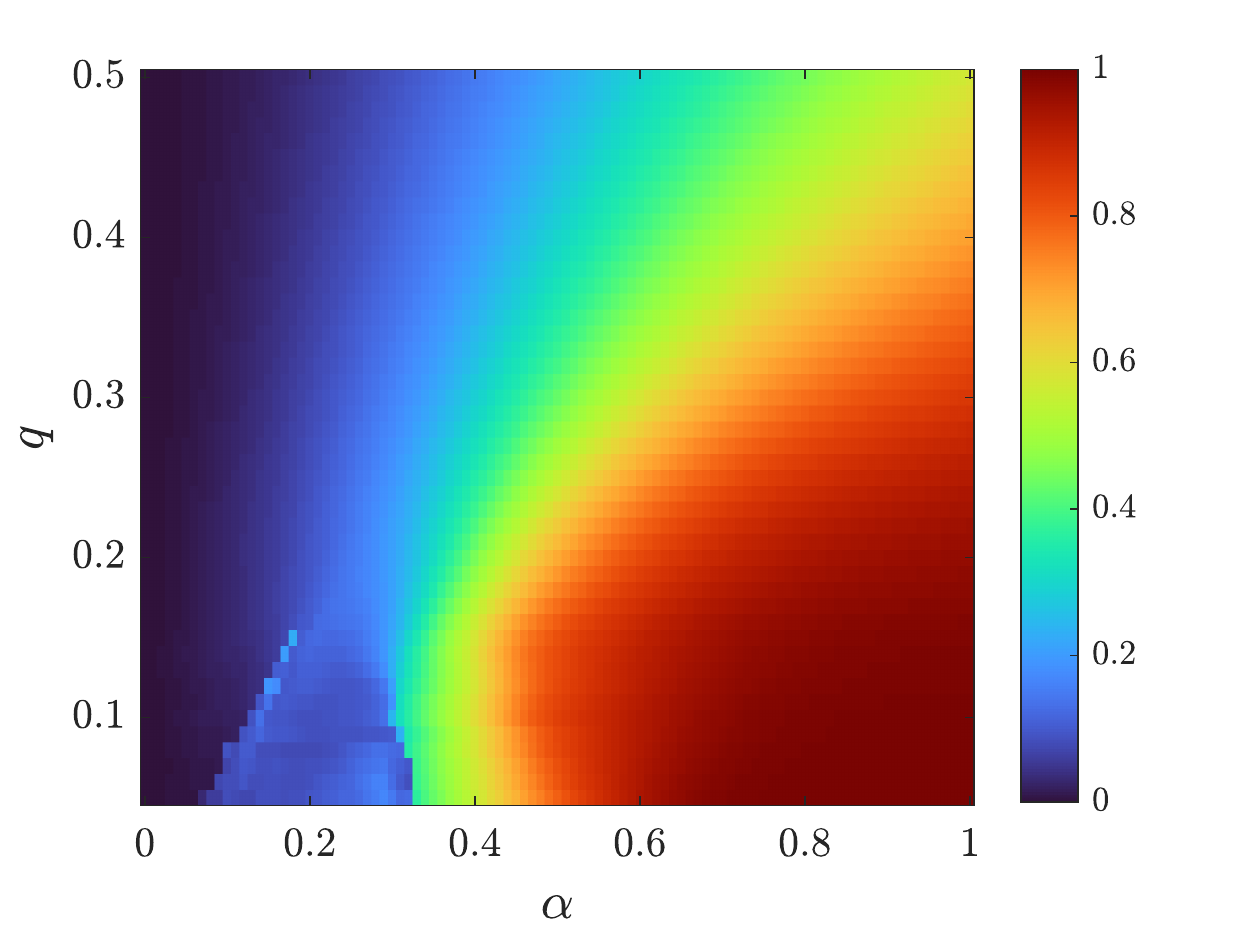} &
\includegraphics[width=6cm]{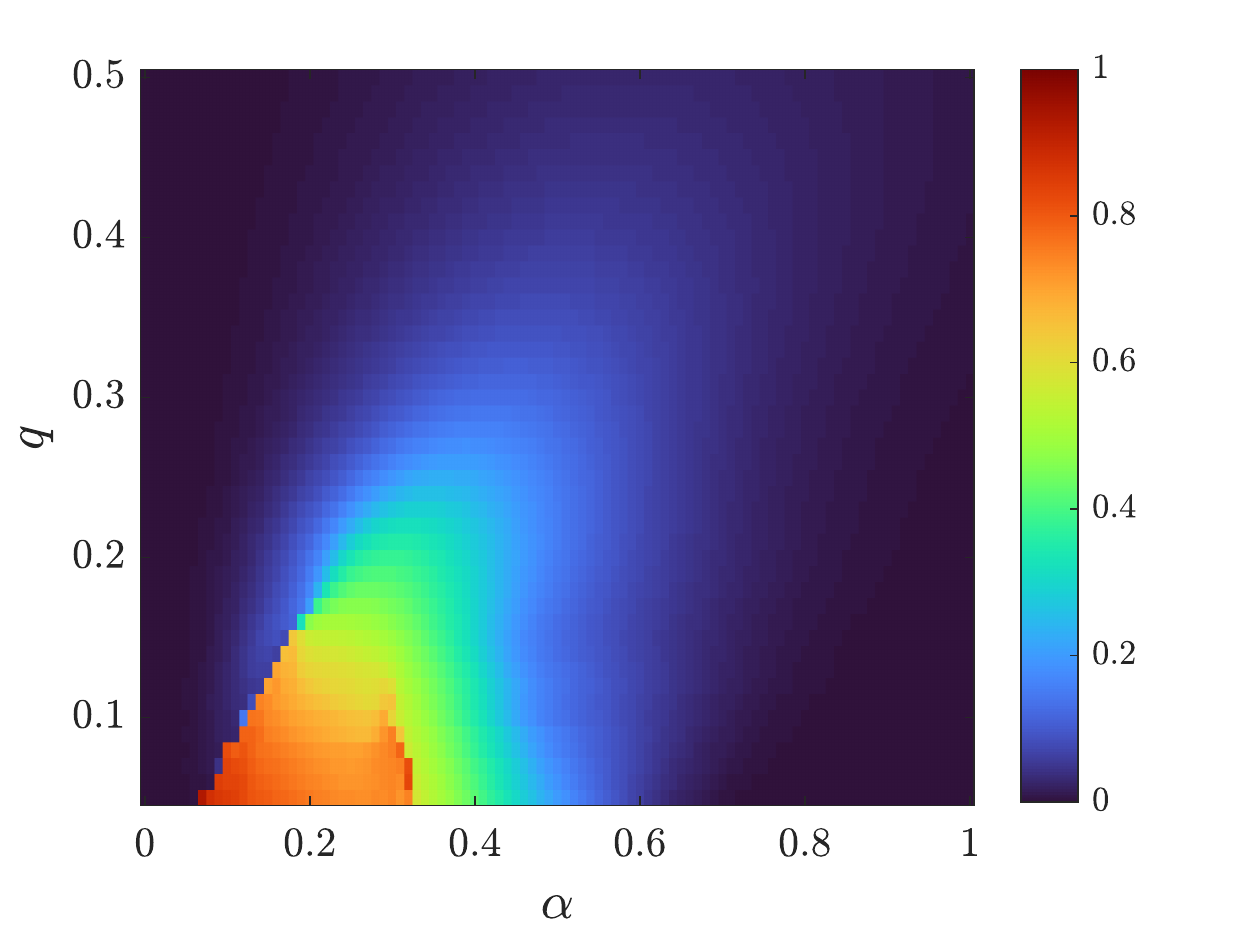} \\
\end{tabular}%
\caption{Attractive impurity. Dependence of $T$ (left panel), $R$ (middle panel) and $L$ (right panel) on $\alpha$ and $q$. In all  simulations, $N=1000$. 
}
\label{fig:TRL_attractive1}
\end{figure}

\begin{figure}
\begin{tabular}{cc}
\includegraphics[width=6cm]{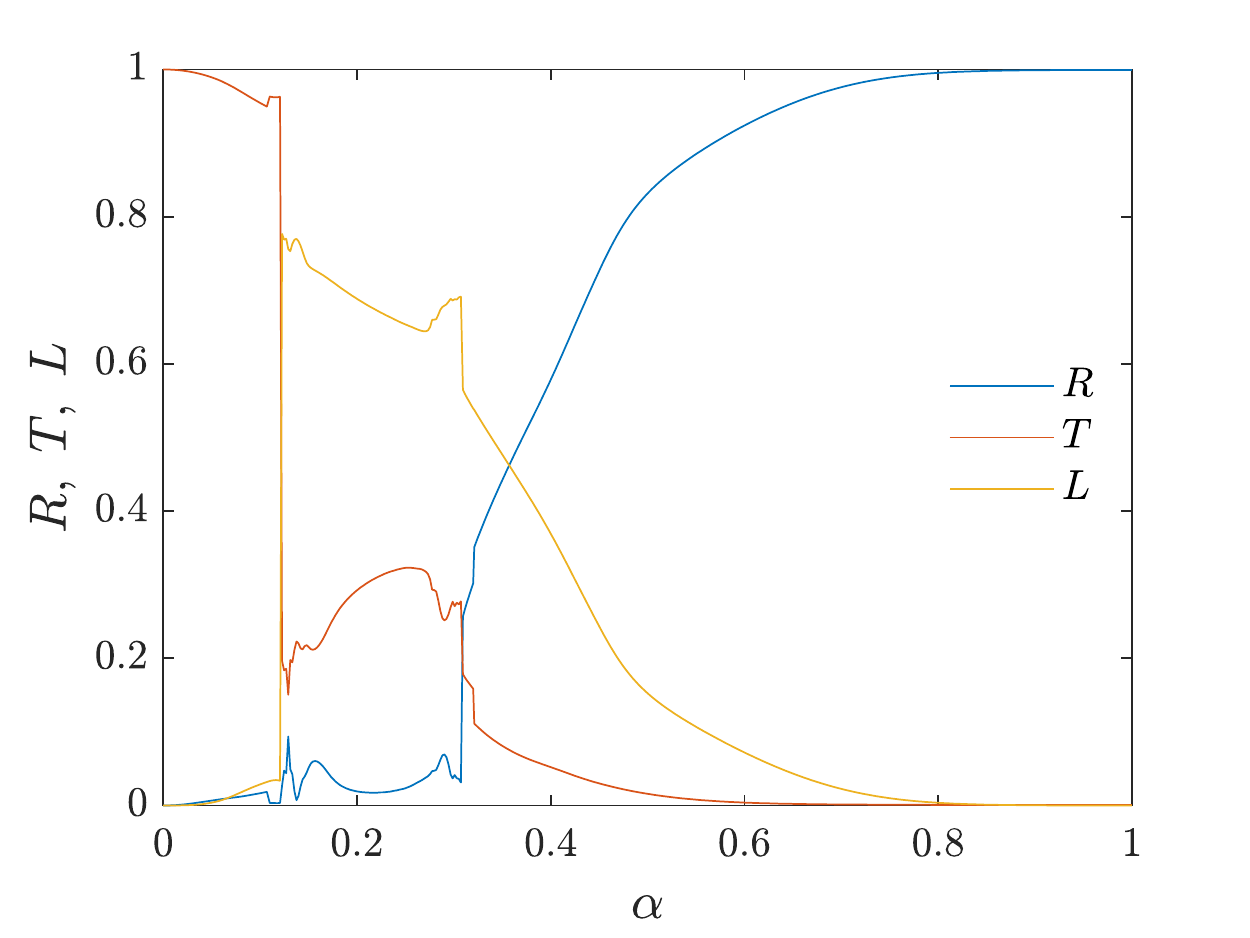} &
\includegraphics[width=6cm]{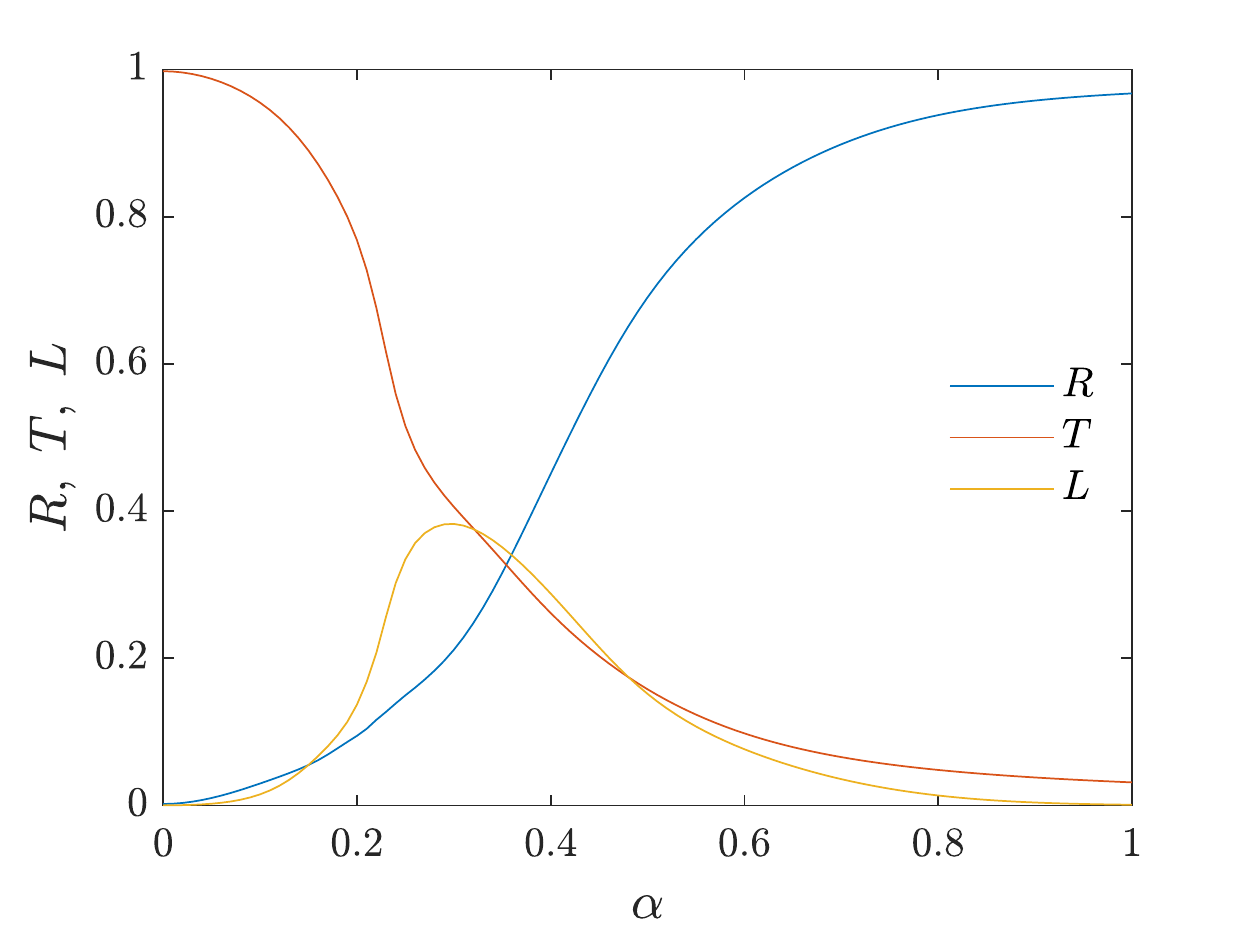} \\
\includegraphics[width=6cm]{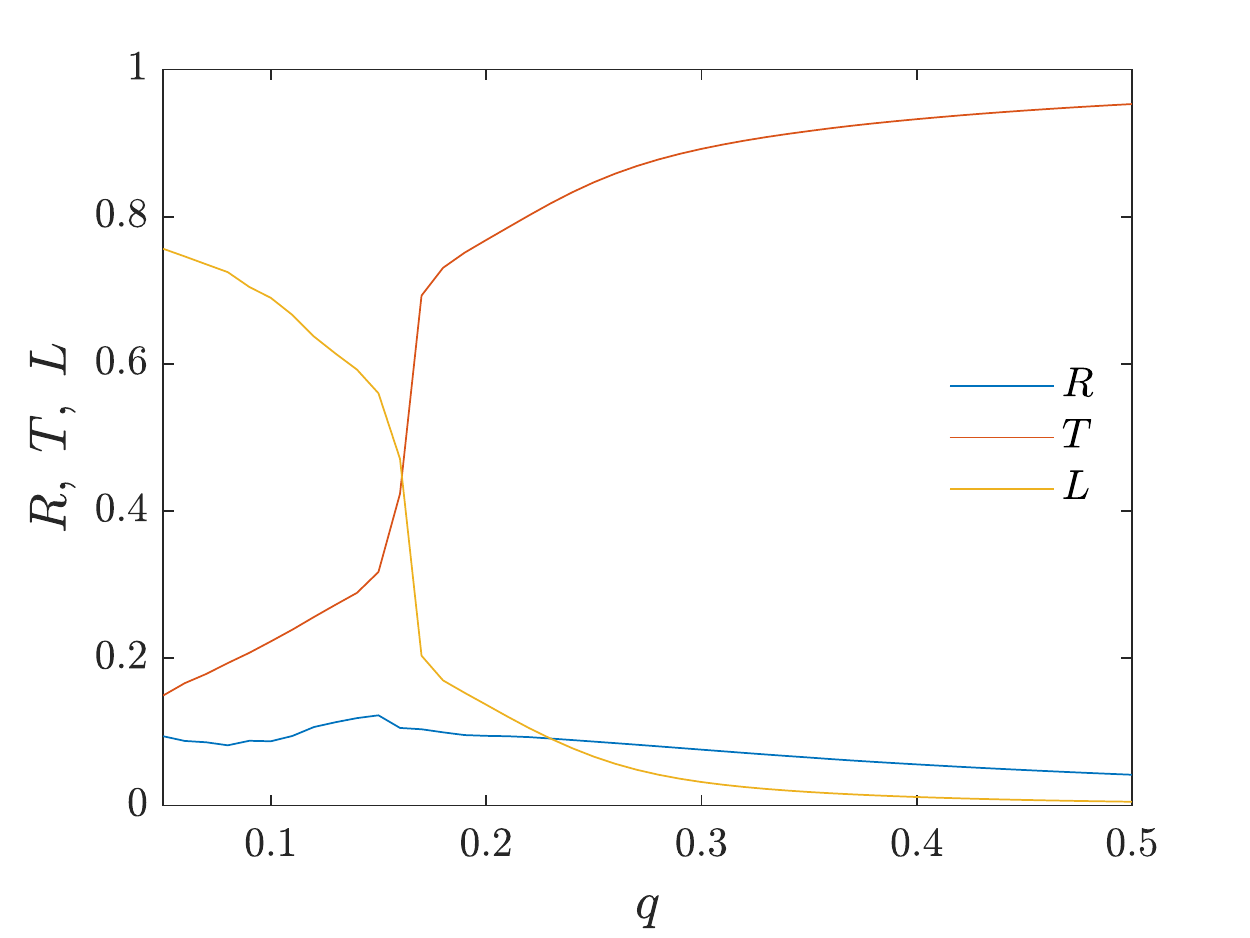} &
\includegraphics[width=6cm]{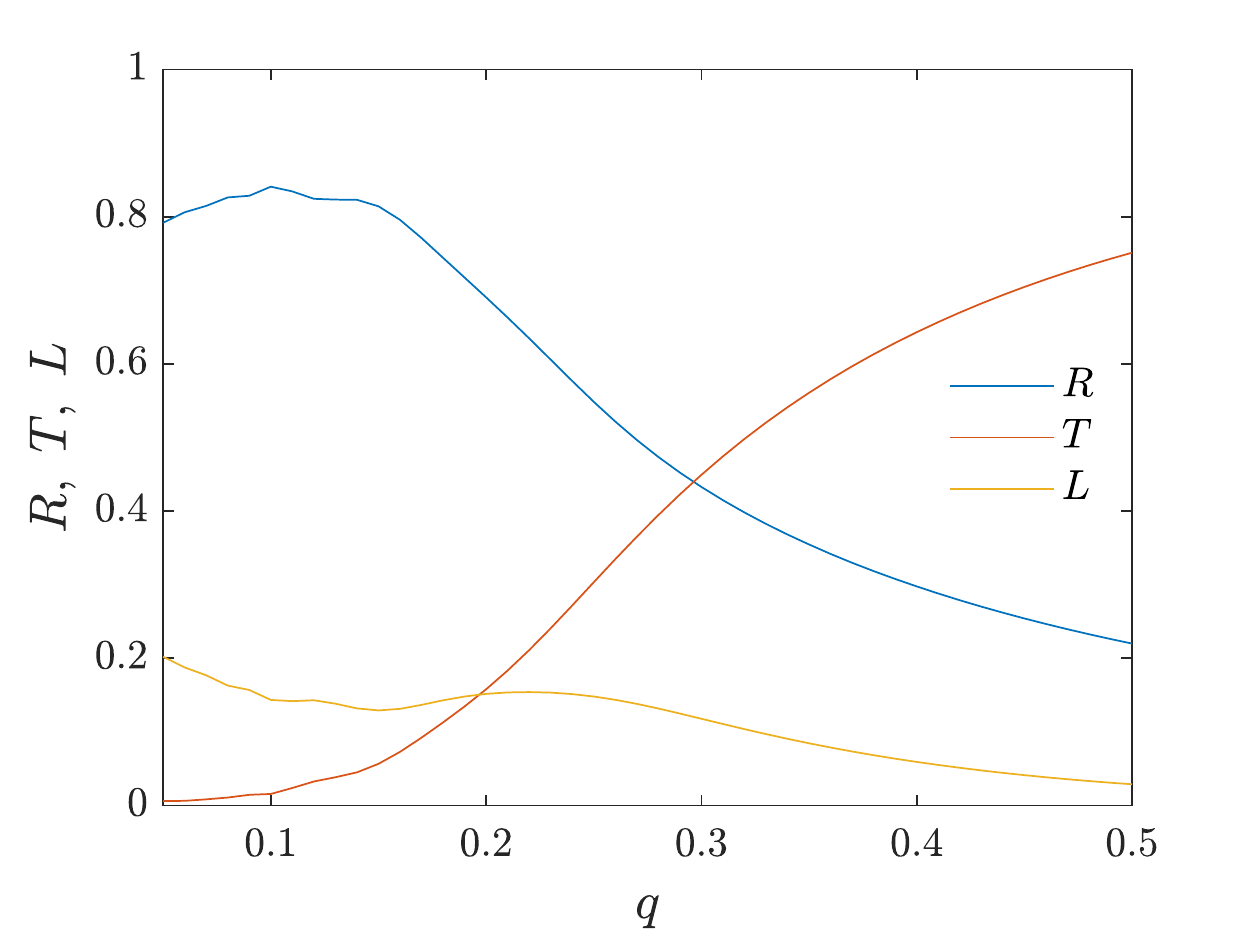} \\
\end{tabular}%
\caption{Dependence of $T$, $R$ and $L$ on $\alpha$, for fixed $q=0.1$ (top left panel) and $q=0.2$ (top right panel), and on $q$, for fixed $\alpha=0.2$ (bottom left panel) and $\alpha=0.5$ (bottom right panel).}
\label{fig:TRL_attractive2}
\end{figure}

Next, we proceed with the case of an attractive impurity, i.e., $\alpha>0$. In this case, the outcome is similar to
that occurring for repulsive impurities, except for the excitation of trapped solitons at the impurity site in some regions of the $(\alpha,q)$ plane. Consequently, the equal splitting scenario is found for $R=T<0.5$, as part of the energy remains trapped. The dependence of $T$, $R$ and $L$ with $\alpha$ and $q$ is depicted in Figs.~\ref{fig:TRL_attractive1} and  Figs.~\ref{fig:TRL_attractive2},  where it is clear
to see that $L$ is nonzero in a localized region of small $q$ and $\alpha$; this is an indication of a nonlinear resonance occurring between the incoming and trapped solitons. We can also observe that for low velocities and small enough $\alpha$, the behaviour of the coefficients is not smooth.

Figure~\ref{fig:simsplit_attractive} shows the evolution of a trapping and splitting process occurring for $\alpha=0.40$ and $q=0.2340$ (corresponding to a velocity $v=0.456$). In this case, we find equal splitting with $R=T=0.4$, and trapping with $L=0.2$; the velocities of the splinters are $0.440$ and $-0.354$. Notice that the trapped soliton has a peaked profile $\sim\exp(a|n|)$, contrary to the incoming and outgoing solitons which preserve the typical of  SDNLS solitons sech / Gaussian shape; this is caused by the peaked shape of nonlinear impurity modes, i.e., solitons centered at the impurity site. On the other hand, the trapped soliton oscillates with a frequency $0.915$; the nonlinear impurity mode with this frequency has been depicted also in Fig.~\ref{fig:simsplit_attractive} with a perfect matching with the trapped soliton, demonstrating that the soliton trapping is actually caused by a nonlinear resonance.

\begin{figure}
\begin{tabular}{ccc}
\includegraphics[width=6cm]{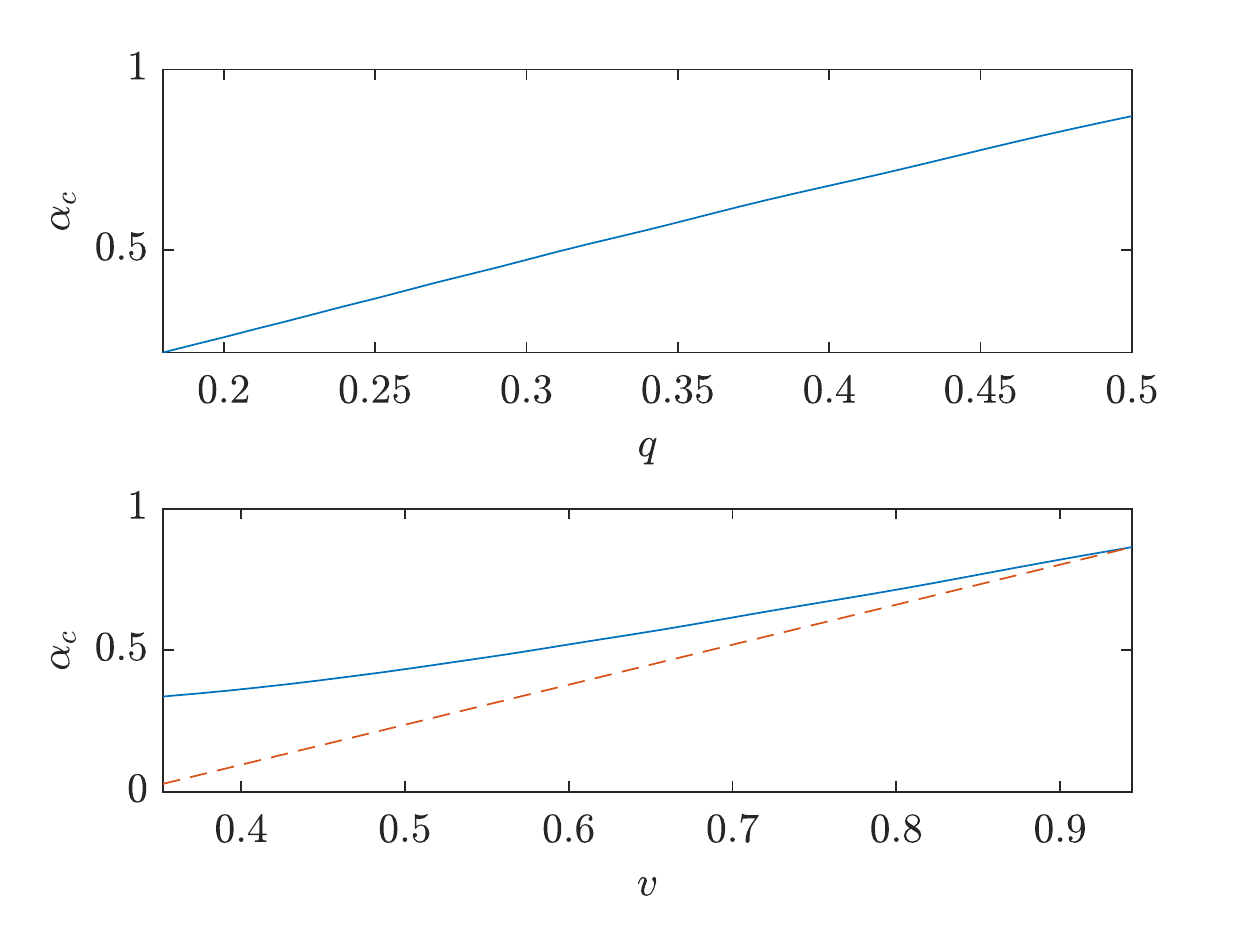} &
\includegraphics[width=6cm]{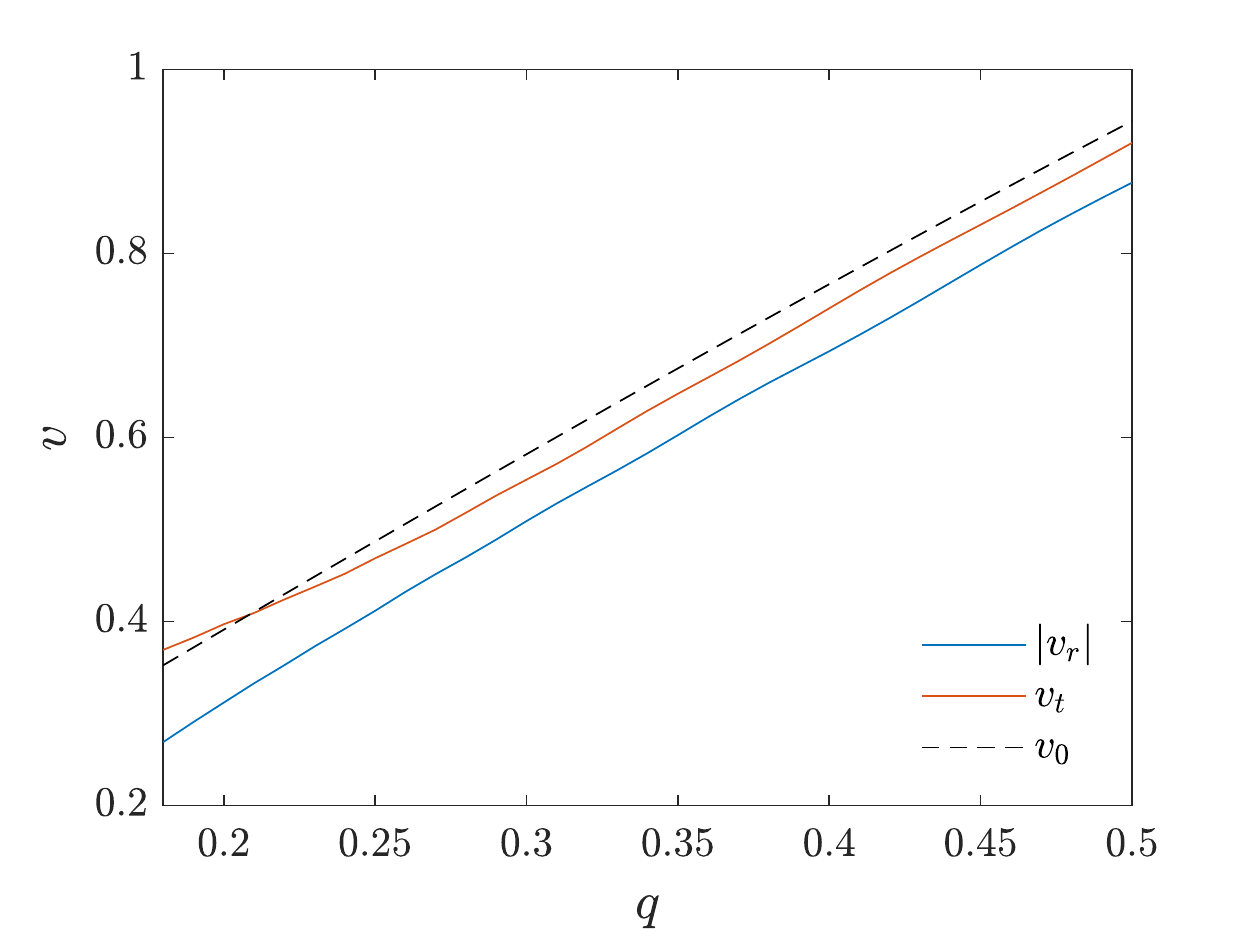} &
\includegraphics[width=6cm]{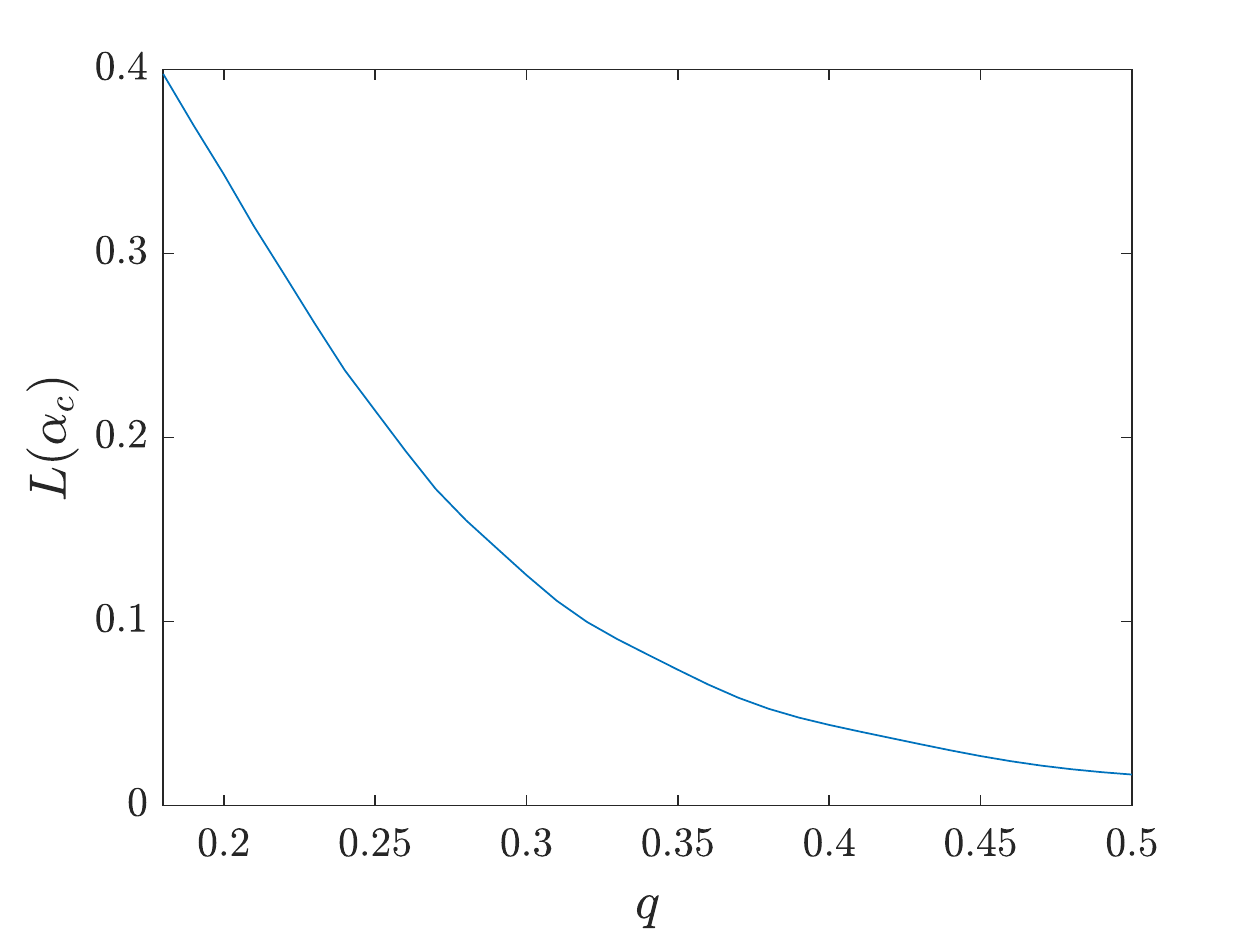}
\end{tabular}%
\caption{Analysis of equal splitting for attractive impurity. Left panel shows the value of $\alpha_c$ versus $q$ and $v$; the red dashed line in the bottom plot corresponds to a line with slope $\sqrt{2C}$, corresponding to the theoretical prediction; middle panel depicts the velocities of the transmitted $v_t$ and reflected $v_r$ splinters together with the velocity of the incoming soliton; right panel displays the trapping coefficient $L$ versus $q$ when $\alpha=\alpha_c$.}
\label{fig:split_attractive}
\end{figure}

Here, it is worth mentioning that, as predicted in Ref.~\cite{Holmer}, the equal splitting takes place for high speed at $\alpha=\alpha_c=v\sqrt{2C}$, similarly to the repulsive impurity case ---see Fig.~\ref{fig:split_attractive}. When comparing to the repulsive case, the velocity of the transmitted soliton is very much closer to that of the incoming soliton, being even higher than the latter for small values of velocities, as can be seen in the middle panel of Fig.~\ref{fig:split_attractive}. This figure also shows the trapping coefficient $L$, for the equal splitting case, featuring a decaying behavior with $q$.

\begin{figure}
\begin{tabular}{cc}
\includegraphics[width=6cm]{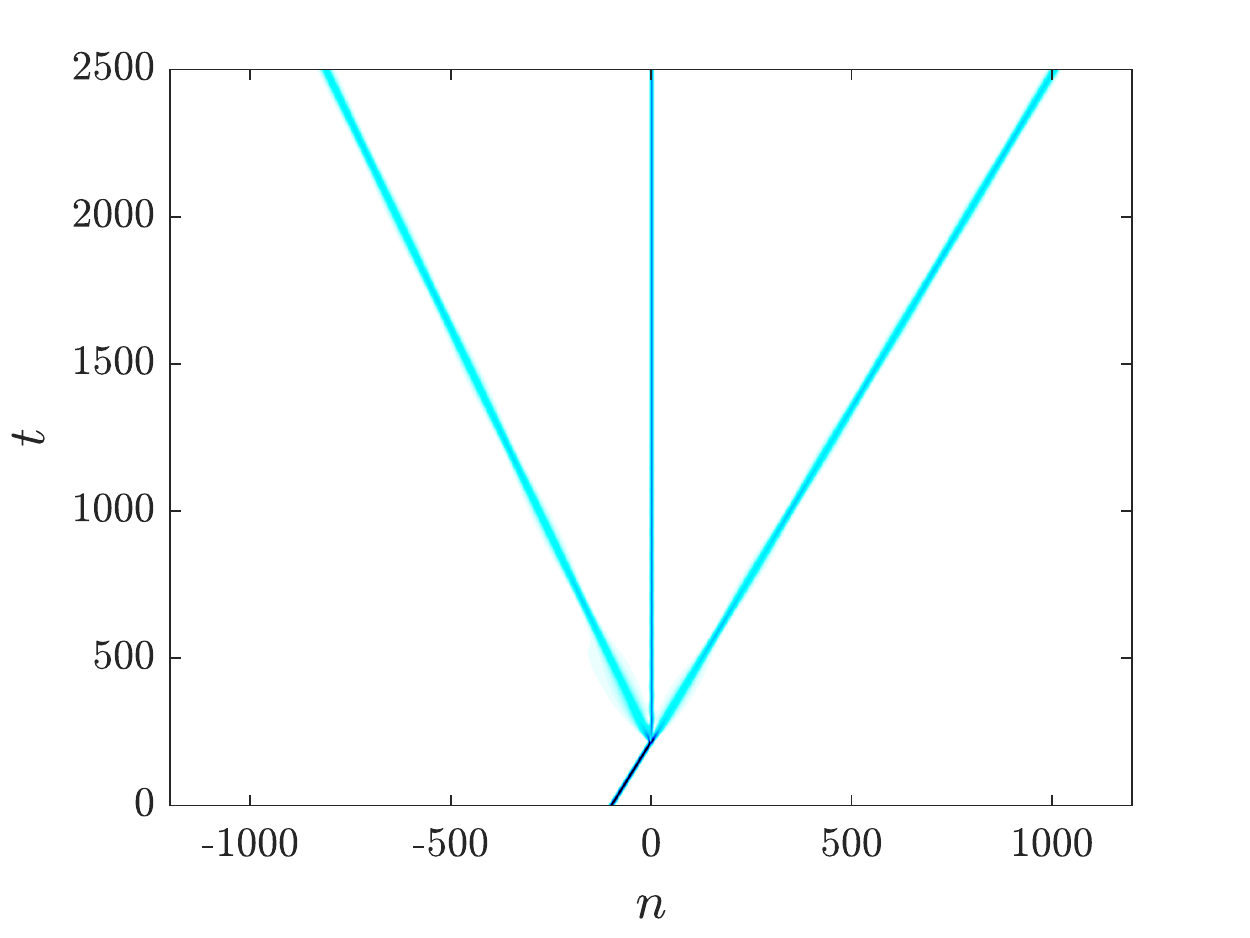} &
\includegraphics[width=6cm]{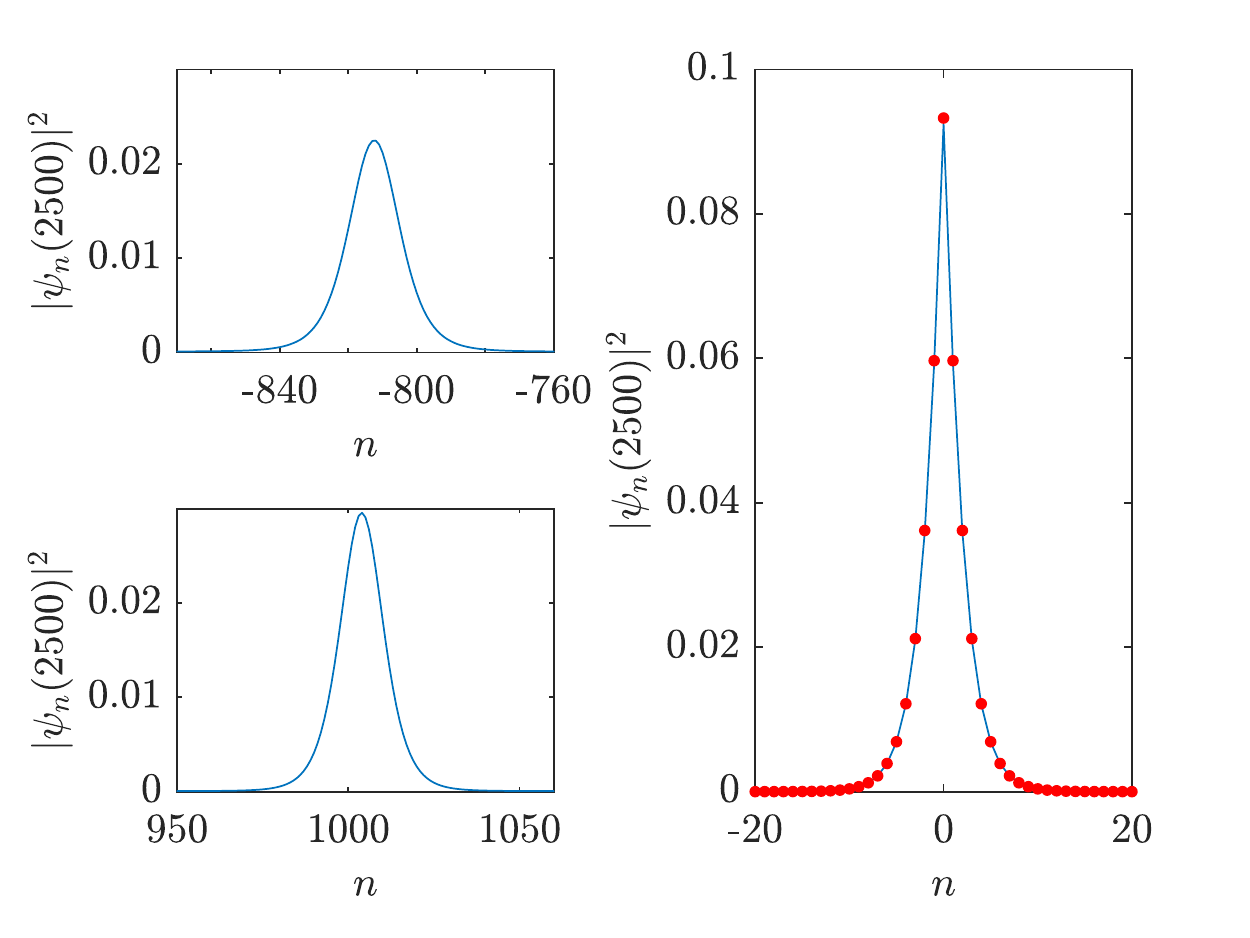} \\
\end{tabular}%
\caption{Evolution of a soliton with $q=0.2340$ interacting with an attractive impurity $\alpha=0.40$ which results in equal splitting and trapping. Left panel shows the space-time diagram of the density $|\psi_n|^2$ and right panel shows the profiles of the splinters and the trapped soliton for a long time ($t=2500$). Notice the red dots overlapping the trapped soliton; it corresponds to a defect mode with the same frequency $\omega=0.915$ as the trapped soliton.}
\label{fig:simsplit_attractive}
\end{figure}

Some of the results presented here are typical for the quantum scattering of nonlinear waves from a localized potential. They are, however, different from those reported in \cite{ricardo} for the cubic DNLS, which are closer to a classical scattering. This is mainly due to the fact that the width of the SDNLS soliton is larger than the effective width of the defect potential, contrary to the case of the soliton of \cite{ricardo}, which is very much narrower. Nevertheless, it should be pointed out that this fact does not preclude the
occurrence of quantum scattering in the cubic DNLS; in fact, if a soliton with $\omega=2.1$ were chosen in the model of \cite{ricardo}, a qualitatively similar behaviour to the reported herein could be observed.

\section{Conclusions} \label{sec:conclusions}

In this work, we have analyzed the interaction of a discrete soliton with an impurity (considered to be either attractive or repulsive) in the saturable nonlinear Schr\"odinger equation. We have found that the scattering process is characterized by typical features of quantum scattering of nonlinear localized waves from narrow inhomogeneities, namely the soliton splits into two splinters and, in the case of attractive impurities, a nonlinear defect mode can be excited resulting in the emergence of trapped solitons. This behaviour is in stark contrast with the observations for narrow solitons interacting with an impurity in the cubic DNLS (see, e.g., the work of Ref.~\cite{ricardo}), where the soliton does not split. Particular attention has been paid to the case of equal splitting and results were successfully compared with the outcome observed for high-speed continuum NLS solitons \cite{Holmer}.

The results presented herein
pave the way for quite interesting future studies. In particular,
the possibility of storing a large amount of energy of SNDLS solitons (compared to cubic ones), specially for frequencies far from the linear modes band, suggests a systematic investigation of
the outcome of the scattering from the impurity, which is expected to be very rich when
weak attractive impurities are considered. In addition, a similar study in the two-dimensional settings, which is impossible in the case of the usual DNLS with the
the cubic nonlinearity but can be realized in the SDNLS equation \cite{vicencio}, may open
avenues for novel unexplored phenomena. These studies are in progress and will be published elsewhere.

\begin{acknowledgments}
We acknowledge support from EU (FEDER program 2014-2020) through both MCIN/AEI/10.13039/501100011033 (under the projects PID2019-108508GB-I00 (FP), PID2019-110430GB-C21 (JCM) and PID2020-112620GB-I00 (JCM)), and Consejer\'{\i}a de Econom\'{\i}a, Conocimiento, Empresas y Universidad de la Junta de Andaluc\'{\i}a (under the projects P18-RT-3480 and US-1380977) (JCM). JFT and AP acknowledge the funding of this study via the project HPC-EUROPA3 (INFRAIA-2016-1-730897), with the support of the EC Research Innovation Action under the H2020 Programme. JFT
gratefully acknowledges computer resources and technical support provided by the Greek Research and Technology Network (GRNET) at HPC-ARIS supercomputer center.
\end{acknowledgments}


\begin{thebibliography}{99}

\bibitem{daux} T. Dauxois and M. Peyrard, {\it Physics of Solitons} (Cambridge University Press, Cambridge, 2006).

\bibitem{kiva} Yu. S. Kivshar and G.P.Agrawal, {\it Optical Solitons: From Fibers to Photonic Crystals} (Academic Press, San Diego, 2003).


\bibitem{reviews_breathers} S. Aubry, Physica D {\bf 103}, 201 (1997); S. Flach and C. R. Willis, Phys. Rep. {\bf 295}, 181 (1998);
P. G. Kevrekidis, K. O. Rasmussen, and A. R. Bishop, Int. J. Mod. Phys. B {\bf 15}, 2833 (2001) ; A. Gorbach and S. Flach, Phys. Rep. {\bf 467}, 1 (2008).

\bibitem{solitons_optics} D.N. Christodoulides, F. Lederer, Y. Silberberg, Nature {\bf 424}, 817 (2003); F. Lederer, G.I. Stegeman, D.N. Christodoulides, G. Assanto, M. Segev, Y. Silberberg, Phys. Rep. {\bf 463}, 1 (2008).

\bibitem{book_panos} P.G. Kevrekidis, {\it The Discrete Nonlinear Schr\"odinger Equation} (Springer-Verlag Heidelberg, 2009).

\bibitem{efremidis} N. K. Efremidis, S. Sears, D. N. Christodoulides, J.W. Fleischer, and M. Segev, Phys. Rev. E {\bf 66}, 046602 (2002).

\bibitem{morandotti} R. Morandotti, U. Peschel, J.S. Aitchison, H.S. Eisenberg and Y. Silberberg, Phys. Rev. Lett. {\bf 83}, 2726 (1999).

\bibitem{luther} W. Kr\'olikowski, B. Luther-Davies and C. Denz, IEEE J. Quantum Electr. {\bf 39}, 3 (2003).

\bibitem{maluckov} L. Had\v{z}ievski, A. Maluckov, M. Stepi\'{c}, D. Kip, Phys. Rev. Lett. {\bf 93}, 033901 (2004).

\bibitem{melvin_prl} T.R.O. Melvin, A.R. Champneys, P.G. Kevrekidis, J. Cuevas, Phys. Rev. Lett. {\bf 97}, 124101 (2006).

\bibitem{vicencio} R.A. Vicencio, M. Johansson, Phys. Rev. E {\bf 73}, 046602 (2006).

\bibitem{kosevich} A.M. Kosevich, Physica D {\bf 41}, 253  (1990).

\bibitem{Cao} X.D. Cao and B.A. Malomed, Phys. Lett. A {\bf 206}, 177 (2005).

\bibitem{Goodman} R.H. Goodman, P.J. Holmes and M.I. Weinstein, Physica D {\bf 192}, 215 (2004).

\bibitem{Holmer} J. Holmer, J. Marzuola and M. Zwoski, J. Nonlinear Sci. {\bf 17}, 349 (2007).

\bibitem{Hulet} J. Cuevas, P.G. Kevrekidis, B.A. Malomed, P. Dyke and R.G. Hulet, New J. Phys. {\bf 15}, 063006 (2013).

\bibitem{Wales} O.J. Wales, A. Rakonjac, T.P. Billam, J.L. Helm, S.A. Gardiner and S.L. Cornish, Commun. Phys. {\bf 3}, 51 (2020).

\bibitem{Polo} J. Polo and V. Ahufinger, Phys. Rev. A {\bf 88}, 053628 (2013).

\bibitem{Sakaguchi} H. Sakaguchi and B.A. Malomed, New J. Phys. {\bf 18}, 025020 (2016).

\bibitem{Ernst} T. Ernst and J. Brand, Phys. Rev. A {\bf 81}, 033614 (2010).

\bibitem{Brand} C. Lee and J. Brand, Europhys. Lett. {\bf 73}, 321 (2006).

\bibitem{Hansen} S.D. Hansen, N. Nygaard and K. M{\o}lmer, Appl. Sci. {\bf 11}, 2294 (2021).

\bibitem{Bishop} R. Scharf and A.R. Bishop, Phys. Rev. A {\bf 43}, 6535 (1991).

\bibitem{Morales} L. Morales-Molina and R. Vicencio, Opt. Lett. {\bf 31}, 966 (2006).

\bibitem{ricardo} F. Palmero, R. Carretero-Gonz\'alez, J. Cuevas, P.G. Kevrekidis and W. Kr\'olikowski, Phys. Rev. E {\bf 77}, 036614 (2008).

\bibitem{Yaounde} J.D. Tchinang Tchameu, A.B. Togueu Motcheyo and C. Tchawoua, Chaos Solitons $\&$ Fractals {\bf 99}, 180 (2017).

\end{thebibliography}
\end{document}